\documentclass[]{aa}  
\usepackage{hyperref}
\hypersetup{
  pdftitle = {AMIGA VIII},
  pdfkeywords = {pdf, hyperref, bookmarks},
  pdfauthor = {Daniel Espada}
}

\usepackage{graphicx}
\def\kms    {\ifmmode{{\rm ~km\,s}^{-1}}\else{~km\,s$^{-1}$}\fi}
\def\arcdeg   {\rm ^{o}}

\def\hi {\ion{H}{I}}

\def\mh2 {M$_{H_2}$}

\usepackage{txfonts}
\usepackage[]{natbib}
\usepackage[usenames]{color}

\bibpunct{(}{)}{;}{a}{}{,} 
\begin{document}

   \title{The AMIGA sample of isolated galaxies}

   \subtitle{VIII. The rate of asymmetric \hi\ profiles in spiral galaxies \dag \thanks{
  Based on observations with the 100-m telescope of the MPIfR (Max-Planck-Institut fuer Radioastronomie) at Effelsberg, GBT under NRAO (the National Radio Astronomy Observatory is a facility of the National Science Foundation operated under cooperative agreement by Associated Universities, Inc.), Arecibo Observatory (National Astronomy and Ionosphere Center, which is operated by Cornell University under a cooperative agreement with the National Science Foundation) and the Nan\c{c}ay Observatory.
\dag\ Full Table~\ref{tab:tests-coeff} is available in electronic form at the CDS via anonymous ftp to {\tt cdsarc.u-strasbg.fr (130.79.128.5)} or via {\tt http://cdsweb.u-strasbg.fr/cgi-bin/qcat?J/A+A/vvv/ppp} and from {\tt http://amiga.iaa.es}. }}

\titlerunning{The AMIGA sample of isolated galaxies.  VIII. \dag}

   \author{D. Espada\inst{1,2} \and  
           L. Verdes-Montenegro\inst{1} \and
           W. K. Huchtmeier\inst{3} \and 
           J. Sulentic, \inst{1}  \and 
           S. Verley \inst{4}   \and 
           S. Leon\inst{5}  \and 
           J. Sabater\inst{1,6} 
        }

   \offprints{D. Espada}

\institute{
 Instituto de Astrof\'{i}sica de Andaluc\'{i}a (IAA-CSIC), Apdo. 3004, 18080 Granada, Spain
 \and
National Astronomical Observatory of Japan (NAOJ), 2-21-1 Osawa, Mitaka, Tokyo 181-8588, Japan; \email{daniel.espada@nao.ac.jp}
 \and
 Max-Planck-Institut fuer Radioastronomie, Postfach 2024, D-53010 Bonn, Germany 
 \and 
 Dept. de F\'{i}sica Te\'{o}rica y del Cosmos, Universidad de Granada, Spain
 \and
Joint ALMA Observatory/ESO, Av. Alonso de C\'{o}rdova 3107, Vitacura, Santiago, Chile
\and 
Institute for Astronomy, University of Edinburgh, Edinburgh EH9 3HJ, UK
 } 

   \date{Received; accepted}

 \abstract
{Measures of the \hi\ properties of a galaxy are among the most sensitive interaction diagnostic
at our disposal. We report here on a study of \hi\ profile asymmetries
(e.g., lopsidedness) in a sample of some of
the most isolated galaxies in the local Universe. This presents us with an
excellent opportunity to quantify the range
of intrinsic \hi\ asymmetries in galaxies (i.e., those not induced by the
environment) and provides us with a zero-point
calibration for evaluating these measurements in less isolated samples.}
{ We aim to characterize the \hi\ profile asymmetries in a sample of
isolated galaxies  and  search for correlations
between \hi\ asymmetry and their environments, as well as their optical and far
infrared (FIR) properties.}
{ We use high signal-to-noise global \hi\ profiles for
galaxies in the AMIGA project (Analysis of the
Interstellar Medium of Isolated GAlaxies, {\tt http://amiga.iaa.es}).
We restrict our study to  $N$ = 166 galaxies
(out of 312) with  accurate measures of the \hi\ shape properties. We
quantify asymmetries using a flux ratio parameter. }
{ The asymmetry parameter distribution of our isolated sample
is well described by a Gaussian model.
The width of the distribution is $\sigma$ = 0.13, and could be even
smaller ($\sigma$ = 0.11) if instrumental errors are reduced.
Only 2\% of our carefully vetted isolated galaxies sample show an
asymmetry in excess of 3$\sigma$.
By using this sample we minimize environmental effects as confirmed by the
lack of correlation between
\hi\ asymmetry and tidal force (one-on-one interactions) and neighbor
galaxy number density. On the
other hand, field galaxy samples show wider distributions and deviate from
a Gaussian curve. As a result
we find higher asymmetry rates ($\sim$ 10--20\%) in such samples. We find
evidence that the spiral arm strength
is inversely correlated with the HI asymmetry. We also find an excess of FIR
luminous galaxies with larger
HI asymmetries that may be spirals associated with hidden accretion
events. }
{ Our sample presents the  smallest fraction of asymmetric \hi\ profiles compared with any other yet studied. The width of the associated asymmetry parameter distribution can help to distinguish the frequency and processes of self-induced HI asymmetries, and serve as a baseline for studying asymmetry rates in other environments.}


   \maketitle

\section{Introduction  }
\label{sec:introduction}

The geometry and kinematics of gaseous disks in galaxies are mainly governed 
by the  gravitational potential of stellar and non-baryonic components.
Any perturbation of the equilibrium caused by either internal or external processes, 
may produce asymmetries in these disks. Atomic gas (\hi ) is
the most extended cold component of the interstellar medium (ISM) and is a sensitive
diagnostic of perturbations. The \hi\ component in spiral galaxies has long been 
known to show both geometrical  and kinematic asymmetries \citep[e.g.,][]{1969Natur.221..531B,1972A&A....17..207H,1973A&A....29..447A,1980MNRAS.193..313B}.
Asymmetries in stellar disks are also common and are traced by optical and 
near-infrared light (the latter less affected by dust extinction). Observations
show that 30\% of galaxies are significantly \emph{lopsided} at near-infrared wavelengths \citep{1994A&A...288..365B,1995ApJ...447...82R,1997ApJ...477..118Z,2005A&A...438..507B}. 
Asymmetries in the stellar component are not necessarily correlated with 
lopsidedness in the gaseous component \citep{2000AJ....120..139K,2004AJ....127.1900W}.
This lack of correlation is not surprising because the neutral hydrogen component in a galaxy
is typically twice as extended as the stellar component \citep[e.g.,][]{1994AAS..107..129B}, and might be perturbed in different ways with respect to the stellar component (e.g., stripping). 
The study of \hi\ properties is thus a better probe of past and/or recent perturbations 
than the stellar counterpart, especially for weak interactions.

Study of global \hi\ velocity profiles of galaxies has proven to be very useful for a
quantification of the frequency and amplitude of disk asymmetries  
\citep[e.g.][]{1994A&A...290L...9R}. While only aperture synthesis can provide 
full 2D information about the \hi\  distribution and kinematics, the 1D profiles provide a 
valuable measurement at a small fraction of the cost in observing time.

Past work suggests that the \hi\  asymmetry properties of galaxies do 
not depend strongly on local environmental conditions. Studies of  \emph{field} and/or 
isolated galaxies suggest that \emph{at least 50~\%} show significant \hi\
profile asymmetries \citep{1994A&A...290L...9R,1998AJ....115...62H}, and 
even higher $\sim$75\% in late-type spiral galaxies \citep{1998AJ....116.1169M}.
Although homogeneous studies of the asymmetry rate in richer environments are rare, it is usually 
believed that they show a comparable rate of asymmetric \hi\ profiles.

The implications of a high asymmetry rate essentially independent of environment is currently 
the subject of debate \citep[for a review see][]{2009PhR...471...75J,2008A&ARv..15..189S}. 
It has been  suggested that the mechanism responsible for producing asymmetric disks 
must be long-lived because high asymmetry rates are observed in samples of field and/or isolated galaxies.
Because the signatures of tidal encounters
are relatively short-lived, lasting only on the order of a dynamical time-scale 
for a wide range in mass ratios, orientations, inclinations, relative velocities and impact parameters  (e.g. \citealt{2005A&A...438..507B}), it cannot be the only agent responsible for 
the high asymmetry rate in different environments.  A number of longer-lived  mechanisms
have been proposed, a) intermittent minor mergers \citep{1996ApJ...460..121W,1997ApJ...477..118Z}, 
b) high-velocity cloud/gas accretion \citep{2005A&A...438..507B,2008A&ARv..15..189S,2008arXiv0810.5130M}, 
c) halo-disk misalignment  \citep{1998ApJ...496L..13L,2001MNRAS.328.1064N} and/or d) internal perturbations 
including sustained long-lived lopsidedness owing to non-circular motions \citep{1980MNRAS.193..313B} 
or global m=1 instabilities \citep{2007MNRAS.382..419S}.

In order to address the relevance of the different proposed mechanisms (internal versus external, 
short versus long-lived) one must first study a sample of well isolated galaxies in the nearby 
Universe ($\lesssim$ 150 Mpc). This approach should minimize any contribution from tidal interactions 
and facilitate the interpretation of results with respect to other samples of galaxies. Reference 
samples used to study the rate of \hi\ asymmetries involve galaxies which, although assumed to 
be field/isolated, usually include a significant population of interacting galaxies \citep[e.g.][]{1994A&A...290L...9R}. 
Note that field galaxies are defined as galaxies not belonging to the cluster environment and a significant number 
of them are  likely to be members of interacting pairs or multiplets  \citep{2006A&A...449..937S}. 
 \citet{1994A&A...290L...9R} find that about half of the nearby field galaxies show asymmetric profiles, 
estimated from a compilation of six \hi\ surveys (1371 profiles were classified). 
It is important to remember that, 1) the overall environmental properties of the samples were not 
assessed meaning that  a significant fraction of galaxies might be environmentally influenced, and 2) asymmetries 
were assessed using qualitative criteria  \citep{1994A&A...290L...9R}.

Statistical studies  of \hi\ asymmetries in large samples of galaxies selected according to a well defined 
isolation criterion and using an objective quantification are rare. The only existing systematic study of 
\hi\ asymmetries in an isolated  sample is that of  \citet{1998AJ....115...62H}, who studied the asymmetry rate 
for $N$ = 104 ($N$ = 78) galaxies that obey a 0.5$\arcdeg$ (1$\arcdeg$ ) projected separation criterion with 
respect to any known companion in the Arecibo General Catalog  (AGC, private database of R. Giovanelli and 
M. P. Haynes), where 0.5$\arcdeg$ corresponds to 175~kpc at a typical velocity of the core sample at V$_r$= 
1500~\kms . About half of the galaxies appear to show significant \hi\ asymmetries. However, only companions 
with a velocity difference $\Delta$V$<$ 400~\kms\ relative to the primary are considered. The AGC is complete 
up to $m$~$\sim$~15.4~mag and/or a diameter of 1\arcmin\ (equivalent to a 
typical linear size of  6~kpc). The velocity criterion could make this sample biased against unbound plunging 
encounters, and owing to the size limit of the catalog, small galaxies would not have been taken into account -- 
hierarchical systems of small galaxies may in principle produce significant asymmetries.

Other studied samples are composed of field galaxies. \citet{1998AJ....116.1169M} studied a sample of $N$ = 30  
moderate to low surface brightness late-type spirals. Their galaxies lie between 2$\arcdeg$ and 6$\arcdeg$ from 
the center of the Fornax cluster or are field galaxies.  They 
 suggest that 77\% of the \hi\ profiles in their sample show a relevant asymmetry.
\citet{2005A&A...438..507B} studied $N$ = 76 galaxies based on the OSUBS Galaxy Survey \citep{2002ApJS..143...73E}, a sample not selected according to any environmental criterion, which hence may have a relevant amount of interacting galaxies. However, it has been recently referred to as a field galaxy sample \citep[e.g.,][]{2009PhR...471...75J}. 
The asymmetry is larger than 10\% for nearly 66\% of the galaxies in this sample.

 In this paper we use a large and complete sample of isolated galaxies \citep{2005A&A...436..443V,2007A&A...462..507L} to 
evaluate the intrinsic distribution of \hi\  asymmetries. Comparisons show that our sample is more 
isolated than those used in previous studies and indeed is representative of the most isolated galaxies 
in the local Universe. The AMIGA project (Analysis of the interstellar Medium of Isolated GAlaxies\footnote{\tt http://amiga.iaa.es}, \citealt{2005A&A...436..443V}) involves vetting and analyzing the properties of galaxies in the Catalogue of Isolated Galaxies ($N$ = 1050 galaxies, CIG, \citealt{1973AISAO...8....3K}), and provides a good starting point for this aim. 
This project includes a refinement of the sample through
1) revision of optical positions \citep{2003A&A...411..391L}, 2) analysis of optical properties and completeness \citep{2005A&A...436..443V}, 3) revised optical  morphologies \citep{2006A&A...449..937S} and 4) reevaluation of the isolation degree \citep{2007A&A...470..505V,2007A&A...472..121V}. We use all these refinements in the present paper. The isolation criterion used for this compilation minimizes the probability of a major interaction within the last  $\sim$ 3~Gyr \citep{2005A&A...436..443V} while quantifying possible minor interactions. 
A multiwavelength characterization of different interstellar medium (ISM) components/phases and of the stellar component has been carried out including a) optical  \citep{2005A&A...436..443V}, b) FIR \citep{2007A&A...462..507L}, c) radio-continuum \citep{2008A&A...485..475L}, and d) H$\alpha$ emission \citep{2007A&A...474...43V}, as well as e) nuclear activity \citep{2008A&A...486...73S,2011A&A...0S}. 

This paper presents an analysis of global (1D) asymmetry measures for $N$ = 312 \hi\ AMIGA galaxies with 
high signal-to-noise (S/N)   spectra (Sect.~\ref{sec:sample}). We cleaned the sample of sources with uncertain 
asymmetry measures, yielding a total of $N$ = 166 galaxies with high reliability data. For this subsample we studied
the \hi\ profile asymmetry rate (Sect.~\ref{sec:lopsidedness}), 
the role of the environment for the rate of lopsided profiles (Sect.~\ref{sec:environment}),
as well as the correlations between \hi\ asymmetry and stellar properties (optical luminosity, morphological type and signs of perturbation), and star formation rate (as traced by FIR luminostity, Sect.~\ref{sec:optical}). Finally we identify the underlying distribution of intrinsic profile asymmetries and discuss the possible origin of asymmetries in gaseous disks (Sect.~\ref{sec:discussion}).

\section{The sample}
\label{sec:sample}

\subsection{Sample selection}
\label{sub:sampleselection}

This paper presents  descriptions of the shape of \hi\ profiles with special emphasis on the degree of asymmetry.  We identified 
the galaxies with the highest $S/N$ \hi\ profiles in the AMIGA project, obtained from observations at the Arecibo, Effelsberg, Nan\c{c}ay and GBT radio-telescopes, as well as from a compilation from archives/literature. The \hi\ spectra are available at {\tt http://amiga.iaa.es}. 
We initially consider for this study those \hi\ spectra with a signal-to-noise $S/N$ $>$ 10 ($S/N$ obtained as the peak flux to rms ratio), in total $N$ = 383 galaxies. 
A $S/N$ $>$ 10 is appropriate to gain a good estimate of the asymmetry \citep{1990A&AS...84...47T}.

In addition we further restricted the sample to galaxies with profiles with high-velocity resolution to total width ratio, $W_{20}$/$\Delta V$ $>$ 10, where $W_{20}$ is the width at a 20\% level and $\Delta V$ is the resolution of the \hi\ profile. 
This criterion is necessary to ensure that the profiles are well sampled. As a side-effect, it partially excludes face-on and low mass/luminosity galaxies.
We excluded sources with evidence of problems related to bad baseline subtraction and/or interference contamination. 
Finally we chose galaxies with recession velocities $V$ $>$ 1500 \kms\  to permit proper evaluation of the isolation.  A prohibitively large region on the sky would be required to assess isolation for galaxies with $V$ $<$ 1500 \ \citep{2007A&A...470..505V,2007A&A...472..121V}. 
In any case, they are all members of the Local Virgo Supercluster. Application of these restrictions yielded a sample of 
$N$ = 312 isolated galaxies with high quality \hi\ profiles. We will refer to this sample as the \emph{\hi\ sample}.
We characterize profile asymmetry in two ways: 1) quality-based on visual inspection of the profiles and 2) quantity-based on an areal asymmetry index $A_{flux~ratio}$ (Sect.~\ref{sec:lopsidedness}). 
For a statistical analysis we performed a further refinement of the sample by considering only those galaxies with lowest uncertainty in their asymmetry parameter. This resulted in a final sample of  $N$ = 166 isolated galaxies, which we call the \emph{\hi\ refined subsample} (Sect.~\ref{sec:cleanedsample}).

\subsection{Basic properties of the samples}
\label{basic}

\begin{figure*}
\includegraphics[width=9cm]{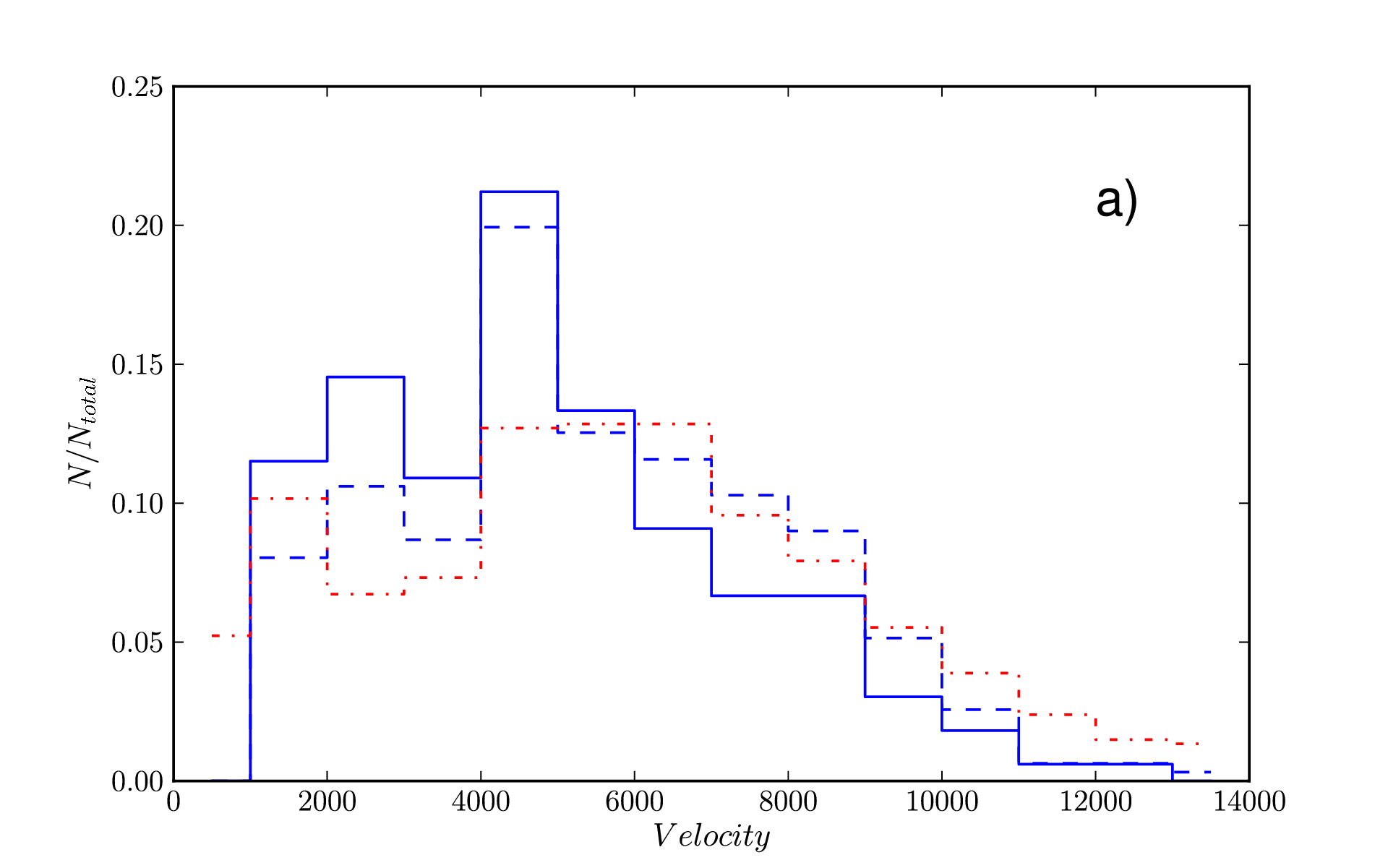}
\includegraphics[width=9cm]{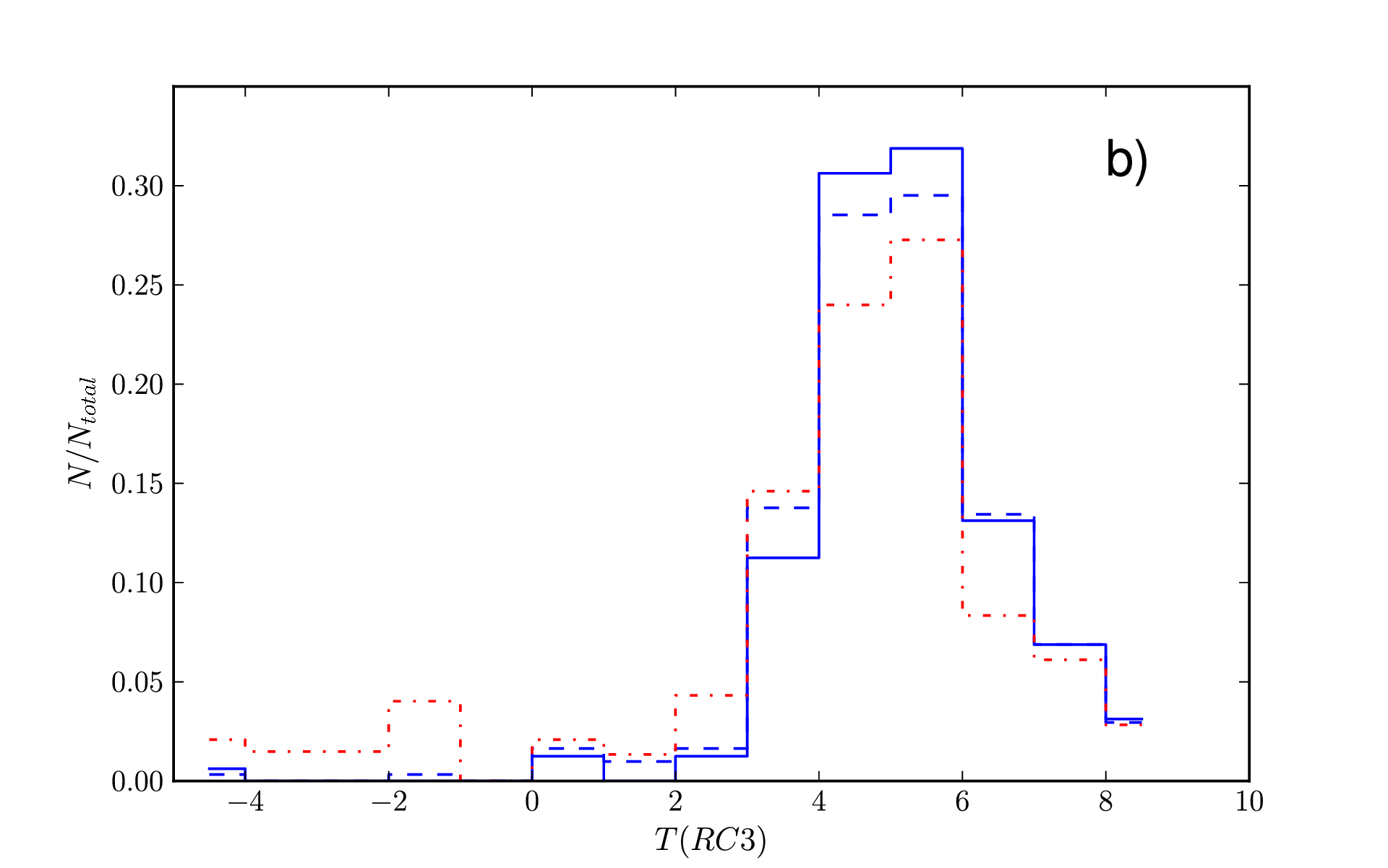}\\
\includegraphics[width=9cm]{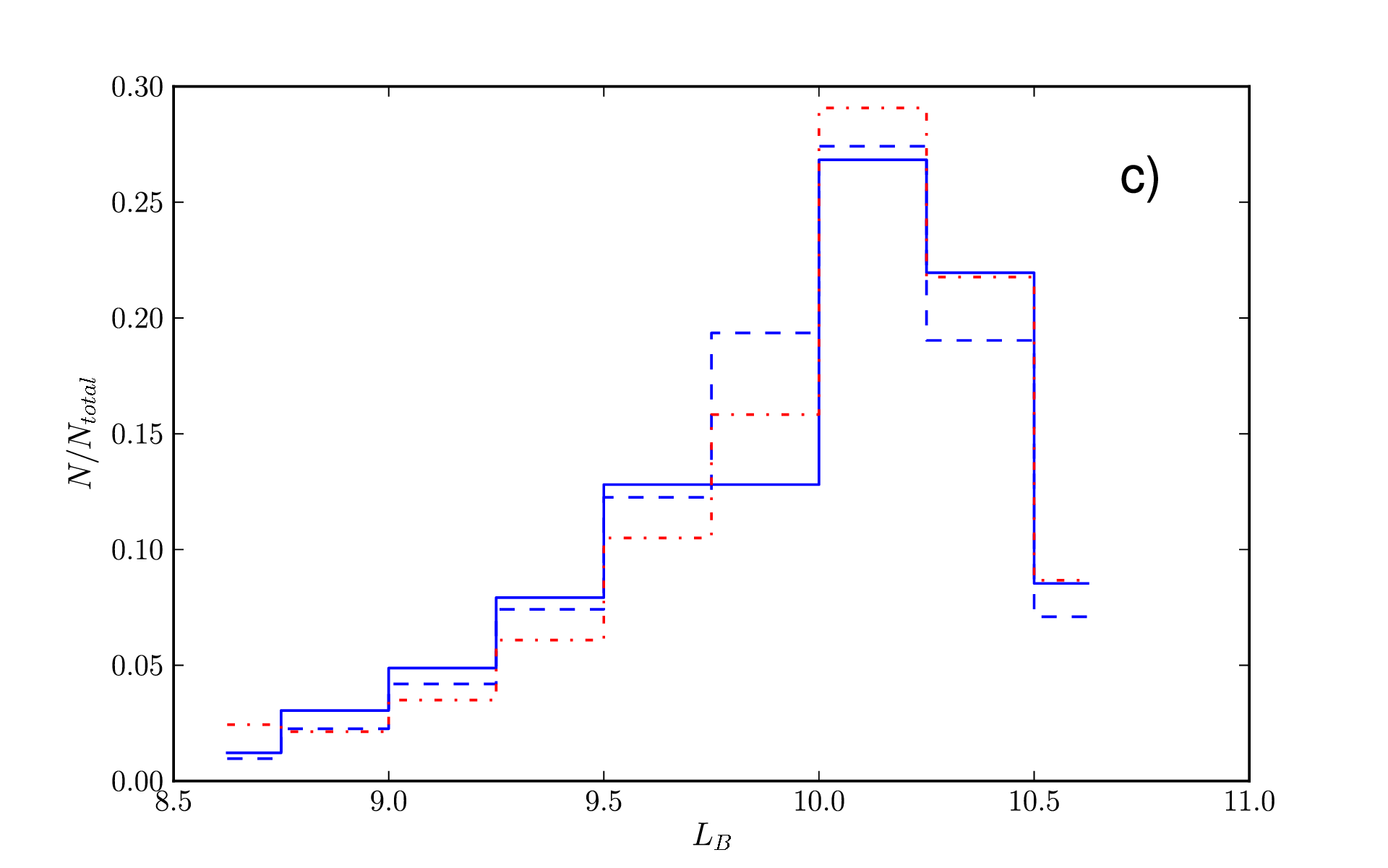}
\includegraphics[width=9cm]{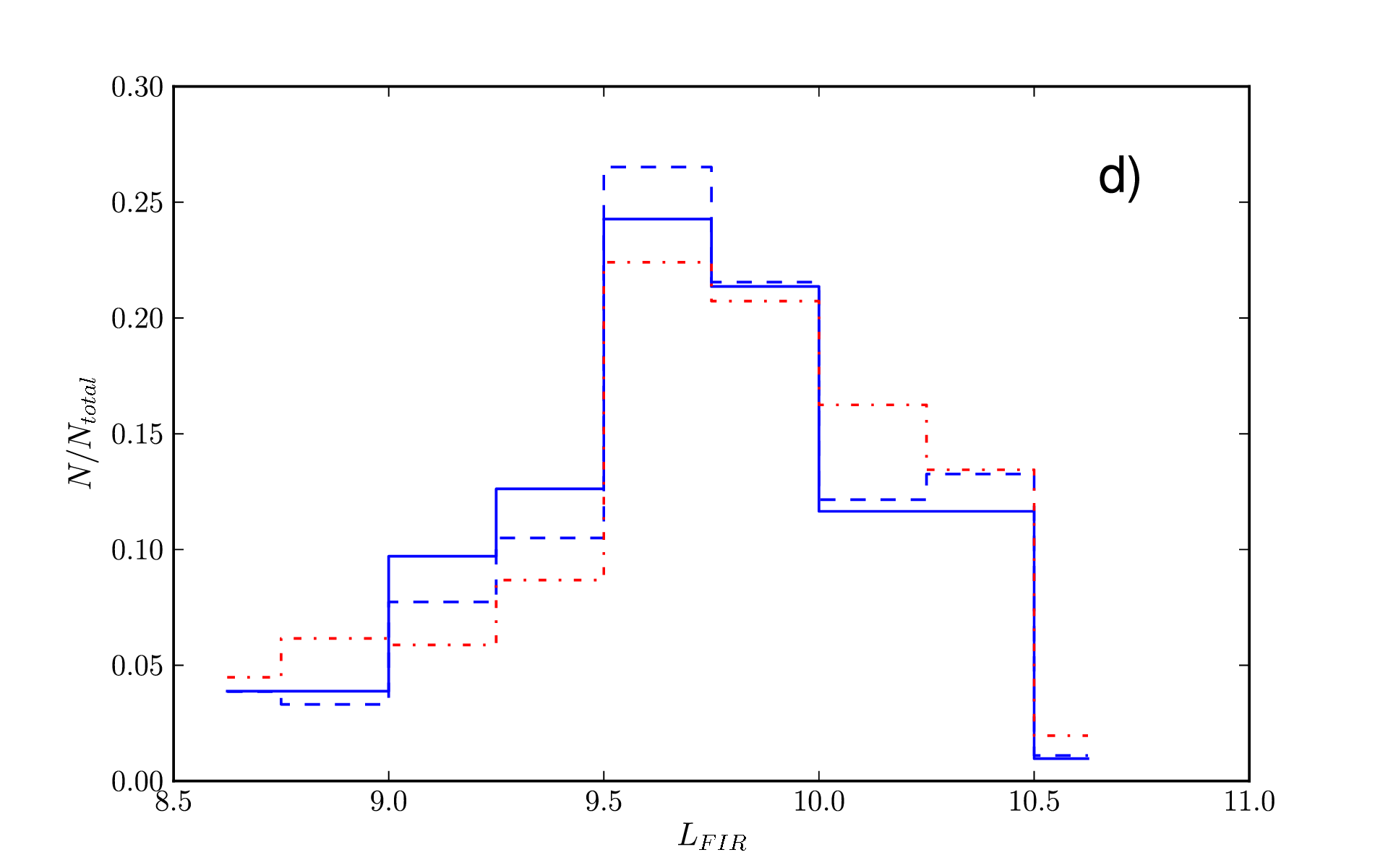}
\caption{Basic properties (normalized distributions) of the \hi\ sample ($N$ = 312 galaxies, blue dashed line), the \hi\ refined subsample ($N$ = 166 galaxies, blue solid  line), as well as the optically complete sample \citep[][red dash-dotted line]{2005A&A...436..443V}: $a)$ velocity  ($V$[\kms]), 
 $b)$ morphology ($T(RC3)$, as in the RC3 catalog),  $c)$ optical luminosity 
 (log($L_B$[$L_\odot$])) and d) FIR luminosity (log($L_{FIR}$[$L_\odot$])).
\label{fig:characterization}}
\end{figure*}

Figure~\ref{fig:characterization} summarizes the basic optical/FIR properties of both the \hi\ sample and the \hi\ refined subsample. The different panels include recession velocity, morphological type as well as optical and FIR luminosity distributions. We compare these distributions with those corresponding to the optically complete sample, which was estimated to be 85--90 \% complete to $m_B$= 15.0~mag,  and is composed of $N$ = 719 CIG galaxies \citep{2005A&A...436..443V}.  The latter sample is appropriate to represent the entire
population of isolated galaxies in the local Universe.

We summarize the dispersions of the basic properties for galaxies included in the different samples as well as the deviations 
between the \hi\ samples and the optically complete sample: 
 
\begin{itemize}
\item $a)$ Radial velocity $V$ (\citealt[][]{2005A&A...436..443V}):  Velocities range over 1dex in the three samples, covering the range 1500~\kms\  $<$ $V$ $<$ 14000~\kms. The \hi\ refined subsample, and to a lower extent the \hi\ sample, is slightly skewed toward lower velocities relative to the optically complete sample. The sample becomes seriously incomplete beyond 9000~\kms.

\item $b)$ Morphology $T(RC3)$  (\citealt[][]{2006A&A...449..937S}): Types are given in the RC3 numerical scale \citep{1991trcb.book.....D}.
The bulk of the CIG sample involves  late-type galaxies in the range 3 $<$ $T$ $<$ 7 (Sb to Sd) with 2/3 of the sample in a very narrow range T=4$\pm$1 (Sb--Sc). Only 14$\%$ of the sample involve early-type systems, which suggests that our sample represents the extreme (low) end of the morphology-density relation. The \hi\ refined subsample contains a higher percentage of late-type galaxies (especially Sb--Sc) than  the optically complete sample. This is not surprising because earlier type galaxies have a systematically lower \hi\ content than later types and 
are especially excluded by selecting those \hi\ profiles with S/N $>$ 10.

\item $c)$ Optical luminosity $L_B$ (\citealt[][]{2005A&A...436..443V}): 
With few exceptions the sample spans a 1~dex luminosity range (9.5 $<$ log($L_B$[$L_\odot$]) $<$ 10.5). Galaxies with 
log($L_B$[$L_\odot$])$<$ 10 are overrepresented in the \hi\ sample (and \hi\ refined subsample) likely because higher luminosity galaxies prefer the high-velocity tail of the sample (Malmquist effect) and often fall below m=15.0. 

\item $d)$ FIR luminosity $L_{FIR}$ (\citealt[][]{2007A&A...462..507L}):
The three distributions are relatively similar to each other.
The shape of the FIR luminosity distribution is flatter than the optical and shows 
a peak near log L$_{FIR}$= 9.6.  The full range covers 2~dex (8.5 $<$ log($L_{FIR}$[$L_\odot$]) $<$ 10.5).
The difference between the optical and FIR luminosity distributions likely reflects the extreme 
FIR "quietness" of our very isolated galaxy sample. Only detections are shown in Figure~\ref{fig:characterization}.
\end{itemize}
 
\section{\hi\ profile shape and quantification of \hi\ lopsidedness }
\label{sec:lopsidedness}

In this section we present a general view of the \hi\ profile shape and two ways of quantifying 
profile lopsidedness. First, we examine the profiles via visual inspection  (see Sect.~\ref{sub:visual}) 
using criteria similar to those employed in the largest \hi\ profile shape study to date, \citet{1994A&A...290L...9R}. 
We then quantify the asymmetry level  in a more objective manner using a numerical parameter (Sect.~\ref{sec:asymmetry_coefficients}) and 
compare visual and numerical descriptions in Sect.~\ref{sec:visualVSquant-Afluxratio}.

\subsection{\hi\ profile shape and visual estimation of lopsidedness}
\label{sub:visual}

\begin{figure*}
\begin{center}
\includegraphics[width=9cm]{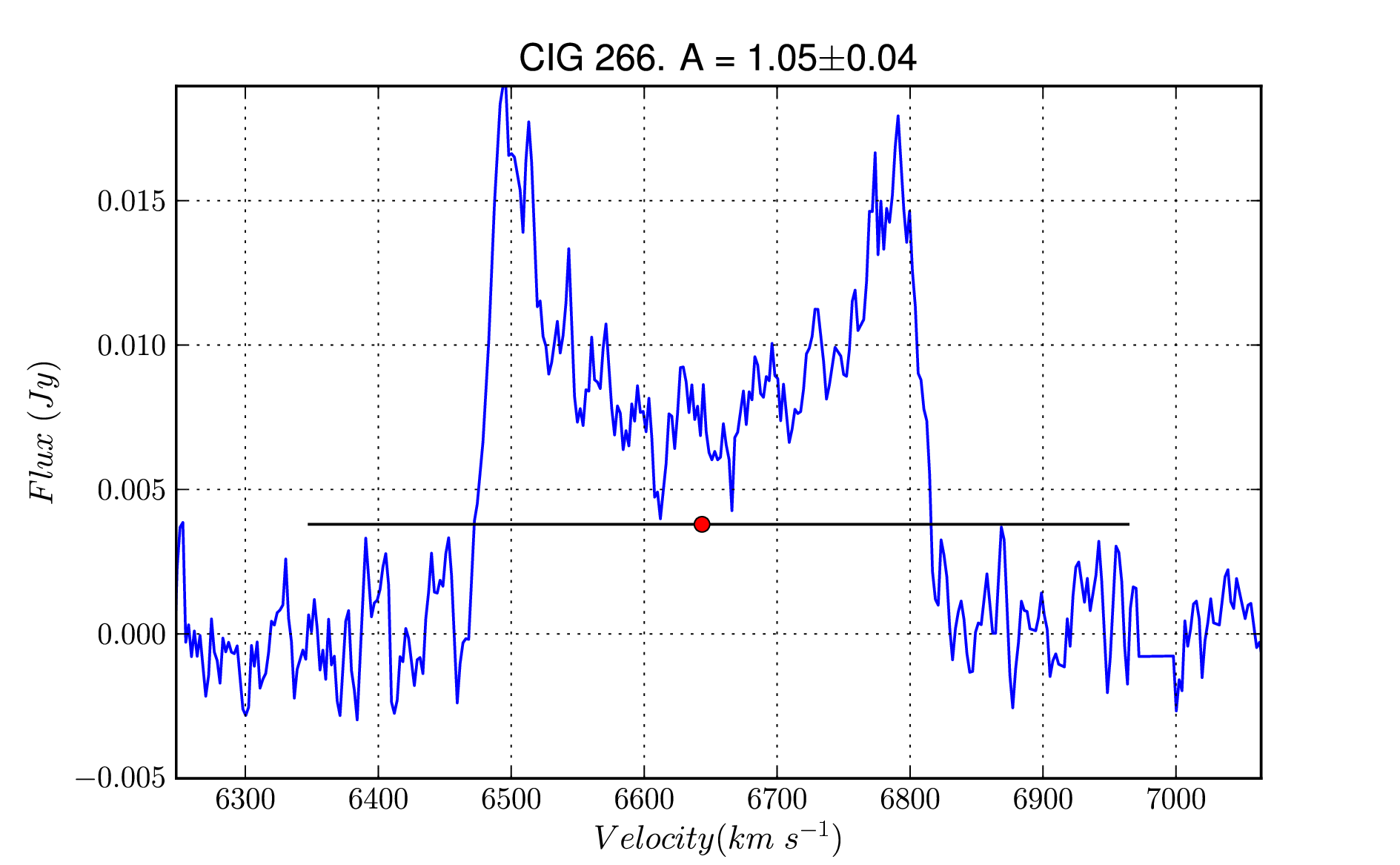}
\end{center}
\caption{Example of a symmetric \hi\ profile: CIG~266, $A_{flux~ratio}$ = 1.05 $\pm$ 0.05.  The points where the horizontal (black) line intersects the profile correspond to the low ($V_l$) and high ($V_h$) velocity ends at a 20\% level with respect to the peak. The derived mean velocity ($V_m$) at a 20\% level is plotted as a (red) point.
   \label{fig:HIProfilesLopsidednessSample1}}
\end{figure*}

\begin{figure*}
\begin{center}
\includegraphics[width=9cm]{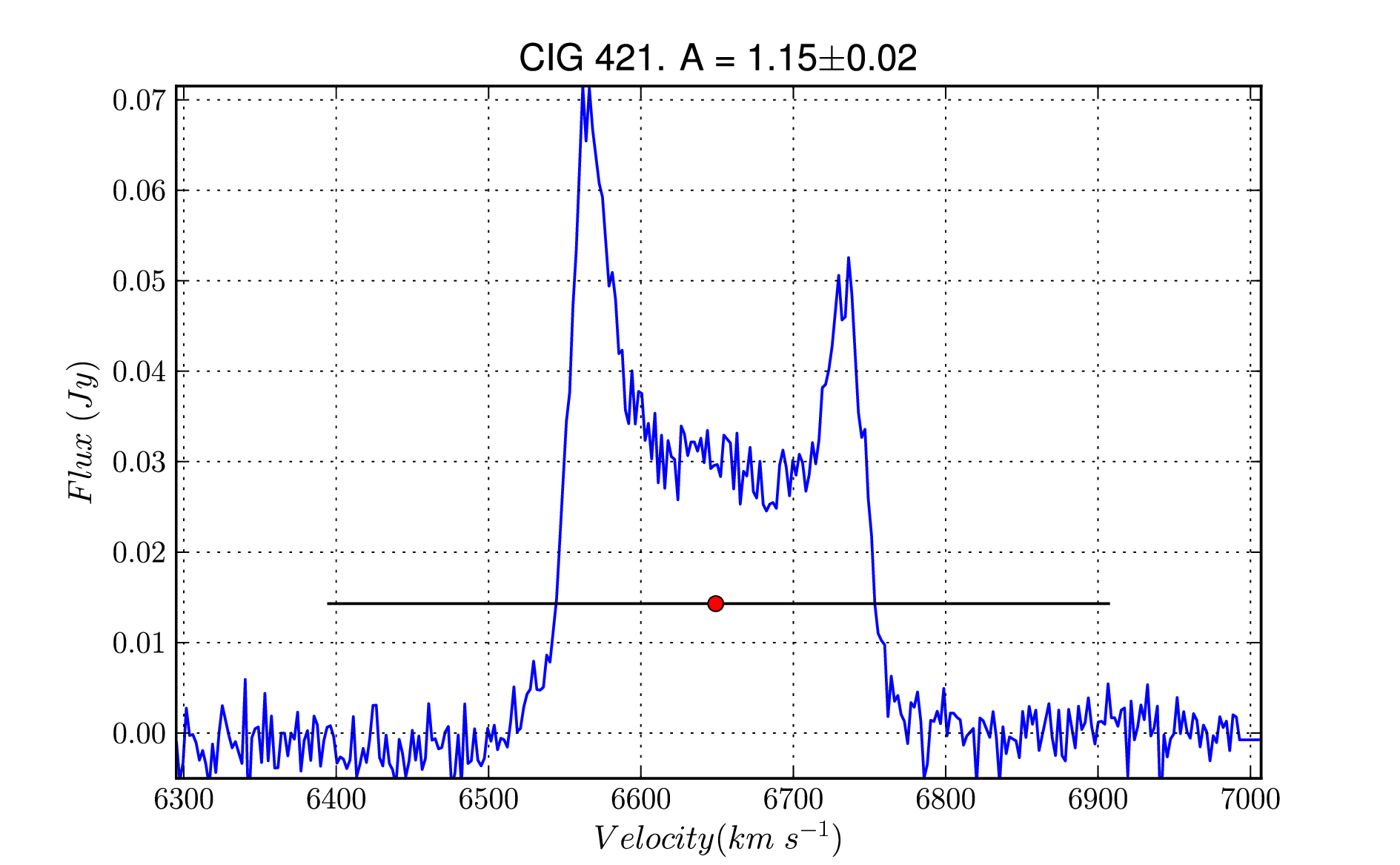}
\end{center}
\caption{Example of a slightly asymmetric \hi\ profile: CIG~421, $A_{flux~ratio}$ $=$ 1.15 $\pm$ 0.03. See description in the caption of Figure~\ref{fig:HIProfilesLopsidednessSample1}.
\label{fig:HIProfilesLopsidednessSample2}}
\end{figure*}

\begin{figure*}
\begin{center}
\includegraphics[width=9cm]{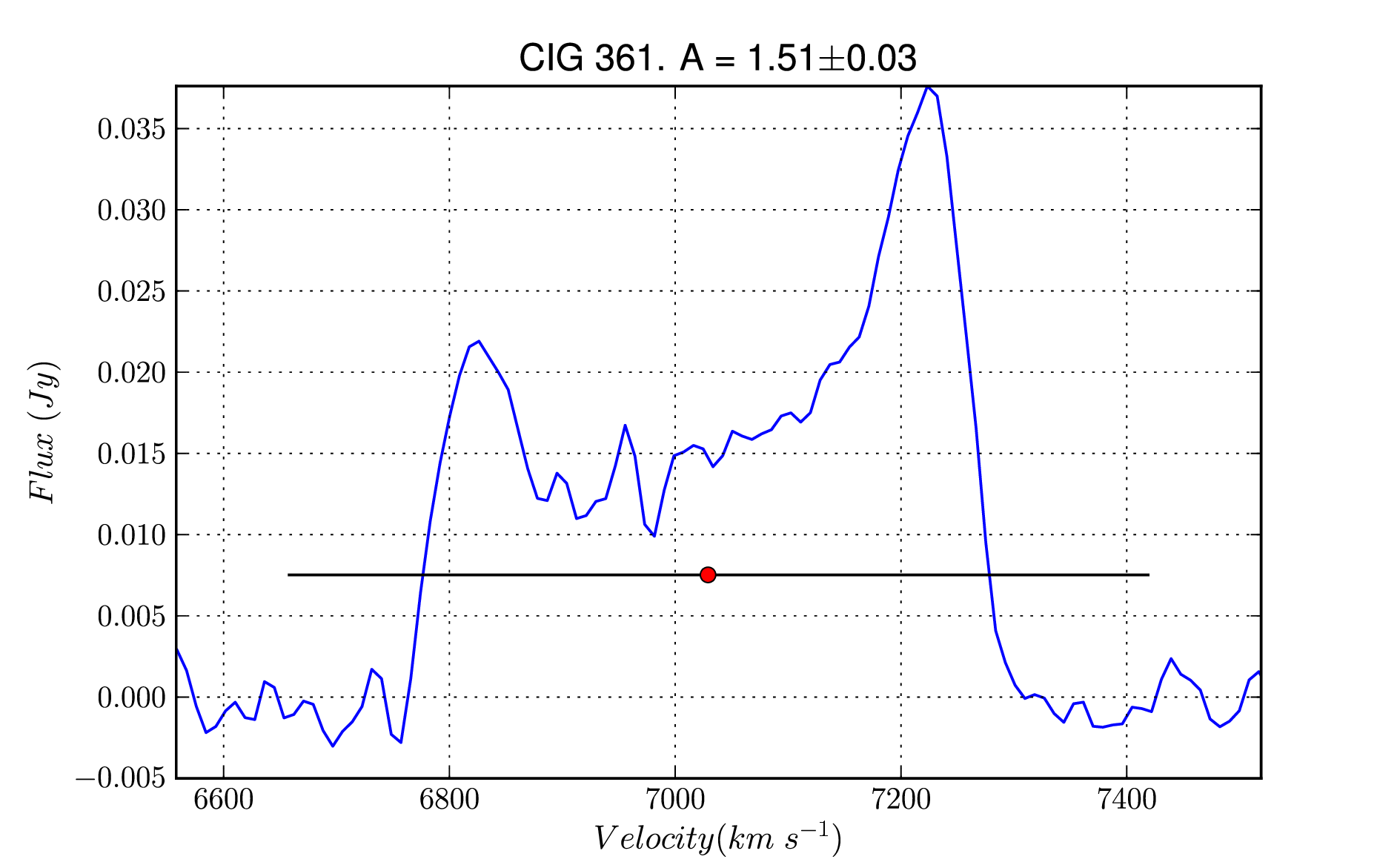}
\end{center}
\caption{ Example of a strongly asymmetric \hi\ profile: CIG~361, $A_{flux~ratio}$ $=$ 1.51 $\pm$ 0.03. See description in the caption of Figure~\ref{fig:HIProfilesLopsidednessSample1}.
\label{fig:HIProfilesLopsidednessSample3}}
\end{figure*}

Visual inspection of our \hi\ sample shows that 88\% show double horns with the rest showing single peaks that 
in most cases involve face-on spiral galaxies. 
We visually classified the profiles in the \hi\ sample ($N$ = 312 galaxies) as $symmetric$, $slightly~asymmetric$, and 
$strongly~asymmetric$, in a similar way as \citet{1994A&A...290L...9R}, who studied a sample of  
$N$ = 1371 spectra (equivalent, respectively, to their notation as  $No$, $Weak$ and $Strong$). 
We find that $N$ = 141 galaxies show symmetric \hi\ profiles (45 \%), $N$ = 126  sligthly asymmetric profiles (40\%), and $N$ = 45  strongly asymmetric profiles (15\%). In order to illustrate this visual classification we  show some examples of symmetric, slightly asymmetric, 
and strongly asymmetric \hi\ profiles in Figures~\ref{fig:HIProfilesLopsidednessSample1}, \ref{fig:HIProfilesLopsidednessSample2}, and \ref{fig:HIProfilesLopsidednessSample3}, respectively.
 Based on the visual classification, our sample appears to show similar average rates as those obtained by \citet{1994A&A...290L...9R}: 47 $\pm$ 5 \%, 34 $\pm$ 6 \%, and 19 $\pm$ 6 \%.

The most common asymmetry found in our sample involves unequal peaks in the double horn profiles. 
The most extreme \hi\ asymmetries occur for 43 galaxies (out of 312) where the peak flux difference 
is approximately larger than 25\%. Only one of them has a peak flux difference larger than 50\% (CIG~317). 
Some galaxies with double peaked profiles show peculiarities: 
CIG~144 shows a central peak stronger than the horns of the double peaked profile,
CIG~858 shows a profile with a peculiar central trough,
CIGs~238, 382, 928, and 1029 show apparent wings (significant excess flux --3$\sigma$-- within a 50 -- 100 \kms\ wide) 
beyond the outer walls of the double-horn profile.
CIG~170 shows an uncommon flat \hi\ profile.
CIG~870 may also have wings that are 50 -- 100 \kms\ wide, although it seems to be a face-on galaxy.
The observed wings may indicate  a projected gas-rich companion or extra-planar motions owing to a nurture event. 

\subsection{Integrated density flux ratio parameter ($A_{flux~ratio}$)}
 \label{sec:asymmetry_coefficients}

A variety of parameters have been used in the literature to quantify the asymmetry  level in \hi\ profiles.
We employ an areal asymmetry index to quantify profile lopsidedness, namely the integrated flux density ratio $A_{flux~ratio}$ (e.g. \citealt{1998AJ....115...62H}, \citealt{2001AJ....121.1358K}), defined as  $A_{flux~ratio}= A_{l/h}$, if $A_{l/h}$ $>$ 1, and $1/A_{l/h}$  otherwise,   where $A_{l/h}$ is the ratio of the areas under the profile at velocities lower (S$_l$) and higher (S$_h$) than the central velocity ($V_m$):

$A_{l/h}=\frac{S_{l}}{S_{h}}=\frac{\int\limits_{V_l}^{V_m}S_{v}dv}{\int\limits^{V_h}_{V_{m}}S_{v}dv}$,

\noindent where $V_l$ and $V_h$ represent the low and high velocities measured at 20\% intensity level 
with respect to the peak. $V_{m}$ is calculated as the mean 
velocity at the same level, {\bf $V_{m}$ = ($V_h$+$V_l$)/2}. Note that $A_{flux~ratio}$ is invariant to the sense of rotation of the galaxy.
We use this parameter since it is the most common asymmetry index that can be found in the literature
and allows us to compare our results with other samples of galaxies (see Sect.~\ref{sub:samples}).
Equivalent definitions are found in the bibliography and can be easily converted to $A_{flux~ratio}$, as e.g.:   $ A=\frac{S_{l}-S_{h}}{S_{l}+S_{h}}$ \citep{1998AJ....116.1169M} and $E1 = 10 \times (1~-~1/A_{flux~ratio})$ \citep{2005A&A...438..507B}.  We indicate the $A_{flux~ratio}$ values in Figure~\ref{fig:HIProfilesLopsidednessSample1}, \ref{fig:HIProfilesLopsidednessSample2}, and \ref{fig:HIProfilesLopsidednessSample3} for the examples of \hi\ profiles 
visually classified as symmetric, CIG~226, slightly asymmetric, CIG~421, and strongly asymmetric, CIG361, which are characterized by $A_{flux~ratio}$ = 1.05, 1.15, and 1.51, respectively. 

\subsubsection{Uncertainties of the $A_{flux~ratio}$}
\label{sec:meanvel}

We estimate the uncertainty of the asymmetry index, $\Delta A_{flux~ratio}$, by taking into account $a)$ the \emph{rms noise} per channel, $b)$ the uncertainty in the calculation of the \emph{mean velocity}, and $c)$ the observational \emph{pointing offsets}:

\begin{itemize}
\item{\emph{a)} Uncertainty owing to the rms of the \hi\ profile ($\Delta A$\emph{($rms$}))}:
owing to the rms of the spectrum, $\sigma$, the uncertainty in $S_l$ is  $ \Delta S_l = \sqrt N_l \sigma R$, where $N_l$ is the number of channels corresponding to  $S_l$ and $R$ is the spectral resolution of the profile. The uncertainty in  $S_h$, $ \Delta S_h$ can be obtained in the same way.
Then $\Delta A_{flux~ratio}$ can be calculated as

$\Delta A_{flux~ratio} =  [(\frac{1}{S_h}\Delta S_l)^2 + (\frac{S_l}{S_h^2} \Delta S_h)^2)]^{1/2}$.

\item{\emph{b)} Uncertainty owing to the measurement of the mean velocity ($\Delta A$\emph{(mean vel)})}: 
since the $A_{flux~ratio}$ is calculated as an areal ratio obtained from the mean velocity, an error in the determination of the latter can induce a wrong measure of $A_{flux~ratio}$. 
An error $\Delta V_m$ in the estimate of the mean velocity $V_m$ produced by limited velocity resolution and/or $S/N$ ratio would artificially increase the asymmetry index of a symmetric profile (i.e. $A_{flux~ratio}$ = 1). Namely
$A_{flux~ratio} = \frac{S_l + \epsilon}{S_h - \epsilon} > 1$. A good estimate of the uncertainty can be obtained from the increase in $A_{flux~ratio}$ for a symmetric profile.
\noindent If the profile were symmetric, then  S$_l$ = S$_h$ = $S$/2, where $A$ is the total area under the profile.  $\epsilon$ can be estimated as $\epsilon$ $\sim$ $h$ $\Delta V_m$, where $h$ is an intensity height scale. The uncertainty of $\Delta V_m$ can be estimated as $\Delta V_m = 4 \frac{\sqrt{R~P}}{S/N} $ \citep{1990A&AS...86..473F}, where $P = (W_{20} -W_{50})/2$, parameter that represents the steepness of the edges of the \hi\ profile, and $W_{20}$ and $W_{50}$ are the widths at 20\% and 50\% with respect to the peak, respectively.  
Owing to the uncertainty in the determination of $V_m$, we can express $ \Delta S_l$ (and $ \Delta S_h$)  as:
$\Delta A_l = \Delta A_h = 4 \frac{\sqrt{R~P}}{S/N} h $, 
\noindent where we estimated that $h = h_{max}/2$, being $h_{max}$ the \hi\ profile  strongest peak.

\item{\emph{c)} Uncertainty owing to pointing offsets ($\Delta A$\emph{(pointing~offset}))}:
 a pointing offset of the antenna with respect to the kinematic center of the galaxy can induce an artificial lopsidedness in the \hi\ profile when the telescope beam is comparable to the size of the galaxy \citep[e.g.][]{1990A&AS...84...47T,2005ApJS..160..149S}.  
 In some cases the observing coordinates were not coincident with the center of the galaxies owing to errors in the positions found in the CIG (e.g. \citealt{2003A&A...411..391L}). 
The expected flux loss (\emph{f}) owing to beam attenuation \emph{and known antenna pointing offsets}  can be calculated from  the optical diameter of the galaxy, beam size, and pointing (\citealt{2005ApJS..160..149S}).
We decomposed the expected flux loss (\emph{f}) into two components, $f$ = $f_{b.a.}$ $f_{p.o.}$, where $f_{b.a.}$  is the flux loss factor arising from beam attenuation and $f_{p.o.}$ is the factor where the contribution of the pointing offset is. The latter is intimately related to the flux loss that contributes to the asymmetry of the \hi\ profile. Note that beam attenuation with no pointing offset causes no asymmetry in the \hi\ profile. 
The difference in the $A_{flux~ratio}$ parameter if the flux loss that contributes to the asymmetry is on the receding or approaching side provides a measure of $\Delta A \emph{(pointing offset)}$. We find among the \hi\ profiles only eight galaxies (out of the 312) that have differences larger than 0.01.
The average is a factor 10 smaller than the contribution from the other two sources of uncertainty. The average is 0.001 and the standard deviation is 0.004. Thus, this source of error is negligible in most cases in our data.
This is a result of the small known pointing offsets in the observations, where the average is equal to 3\arcsec\ and the standard deviation is 3\arcsec .

\end{itemize}

We added these sources of uncertainty in quadrature to estimate the net uncertainty 
in $A_{flux~ratio}$, $\Delta A_{flux~ratio}$.

\subsubsection{Other possible sources of uncertainty}
\label{subsec:otheruncertainties}
In addition to these sources of uncertainty, there are other effects that can induce an artificial asymmetry on the \hi\ profiles. These include the effect of \emph{random} pointing offsets and baseline fitting.

Flux loss due to this effect must be somewhere between $\lesssim$ 1\% (GBT, \citealt{1998AJ....115...62H}) and 5\% (Arecibo circular feed, \citealt{1984AJ.....89..758H}). If the profile were initially symmetric, then the induced asymmetry parameter by this effect would be in the range $A_{flux~ratio}$ $<$ 1.02 -- 1.11, if all flux loss is located either in the receding or approaching sides. Because we have data from different telescopes, our situation is probably  intermediate between both cases. We assume that the resulting $A_{flux~ratio}$ distribution of a sample with symmetric \hi\ profiles observed under similar conditions as our sample is likely well represented by a half-Gaussian with a  $\sigma$ = 0.04.

The baseline fitting process can also produce artificial asymmetries in the \hi\ profiles \citep{1998AJ....115...62H}. \citet{1998AJ....115...62H} indicate that different order fits show flux differences of about 3\%. As a result, the asymmetry parameter for symmetric \hi\ profiles can be altered up to $A_{flux~ratio}$ = 1.06. 
A half-Gaussian curve with $\sigma$ = 0.02 would mimic this effect well.

Given the random nature of these two effects, we cannot estimate their values individually, but their overall effect is taken into account in Sect.~\ref{sec:discussion} to discuss the actual shape of the $A_{flux~ratio}$ distribution in isolated galaxies because they can broaden the resulting distribution.

\subsubsection{Presentation of the asymmetry data}
\label{subsec:presentation}
We list in Table~\ref{tab:tests-coeff} the following information: 
\begin{itemize}

\item \emph{1)} CIG number;
\item \emph{2)} Visual classification: $0$ = symmetric, $1$ = slightly asymmetric, and $2$ = strongly asymmetric (Sect.~\ref{sub:visual}),
\item \emph{3)} $A_{flux~ratio}$,  the asymmetry parameter  (Sect.~\ref{sec:asymmetry_coefficients}), 
\item  \emph{4)} $\Delta A (rms)$, the uncertainty in $A_{flux~ratio}$ owing to the rms of the \hi\ profile (Sect.~\ref{sec:meanvel}),
\item  \emph{5)} $\Delta A (mean~vel)$, the uncertainty in $A_{flux~ratio}$ owing to the mean velocity (Sect.~\ref{sec:meanvel}),
 and 
 \item \emph{6)} $\Delta$$A_{flux~ratio}$, the global uncertainty in $A_{flux~ratio}$, including the small contribution from $\Delta$$A(pointing~offset)$ (Sect.~\ref{sec:meanvel}).
\end{itemize}

\begin{figure*}
\begin{center}
\includegraphics[width=12cm]{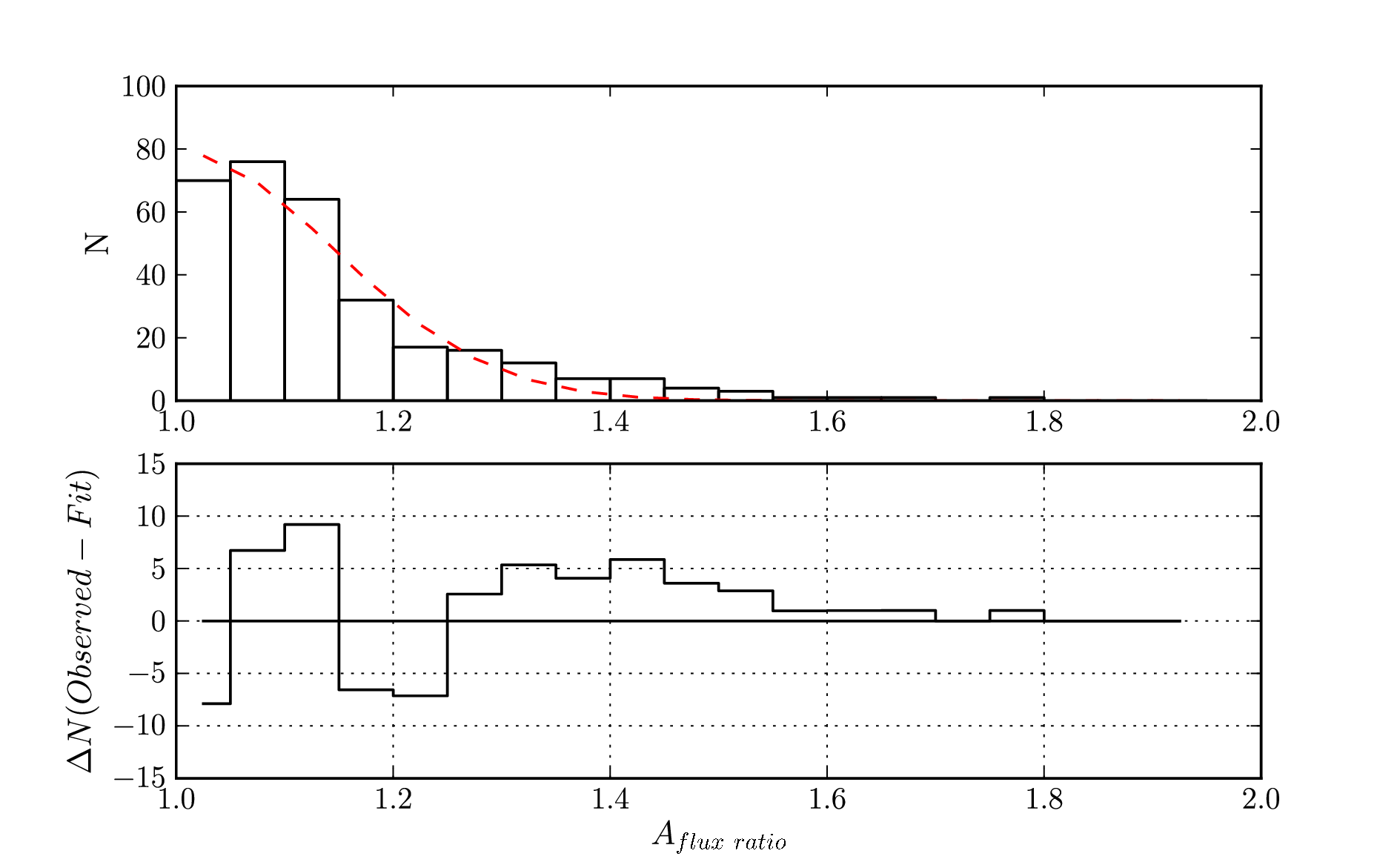}
\caption{$Upper~panel)$ The $A_{flux~ratio}$ distribution (solid line histogram) of the \hi\ sample ($N$ = 312) and its best half-Gaussian fit (dashed line).  $Lower~panel)$ The residual of the half-Gaussian fit to the observed $A_{flux~ratio}$ distribution. 
\label{fig:lops-visual-sample}}
\end{center}
\end{figure*}

The $A_{flux~ratio}$ distribution is shown in Figure~\ref{fig:lops-visual-sample}. The best half-Gaussian fit\footnote{We fitted the parameters $A$, $\mu$ and $\sigma$ in a half-Gaussian curve defined as $ | A~\rm exp(\frac{-(x-\mu)^2}{2. \sigma^2})|$}  to the asymmetry parameter distribution is characterized by a $\sigma$ = 0.15.
However, this half-Gaussian fit is not able to reproduce the $A_{flux~ratio}$ distribution both at the high and low ends. First there is an excess of high values of $A_{flux~ratio}$ with respect to the Gaussian curve, and second, the peak of the distribution is too flat for $A_{flux~ratio}$ $<$ 1.15.

\begin{figure*}
\begin{center}
\includegraphics[width=8.9cm]{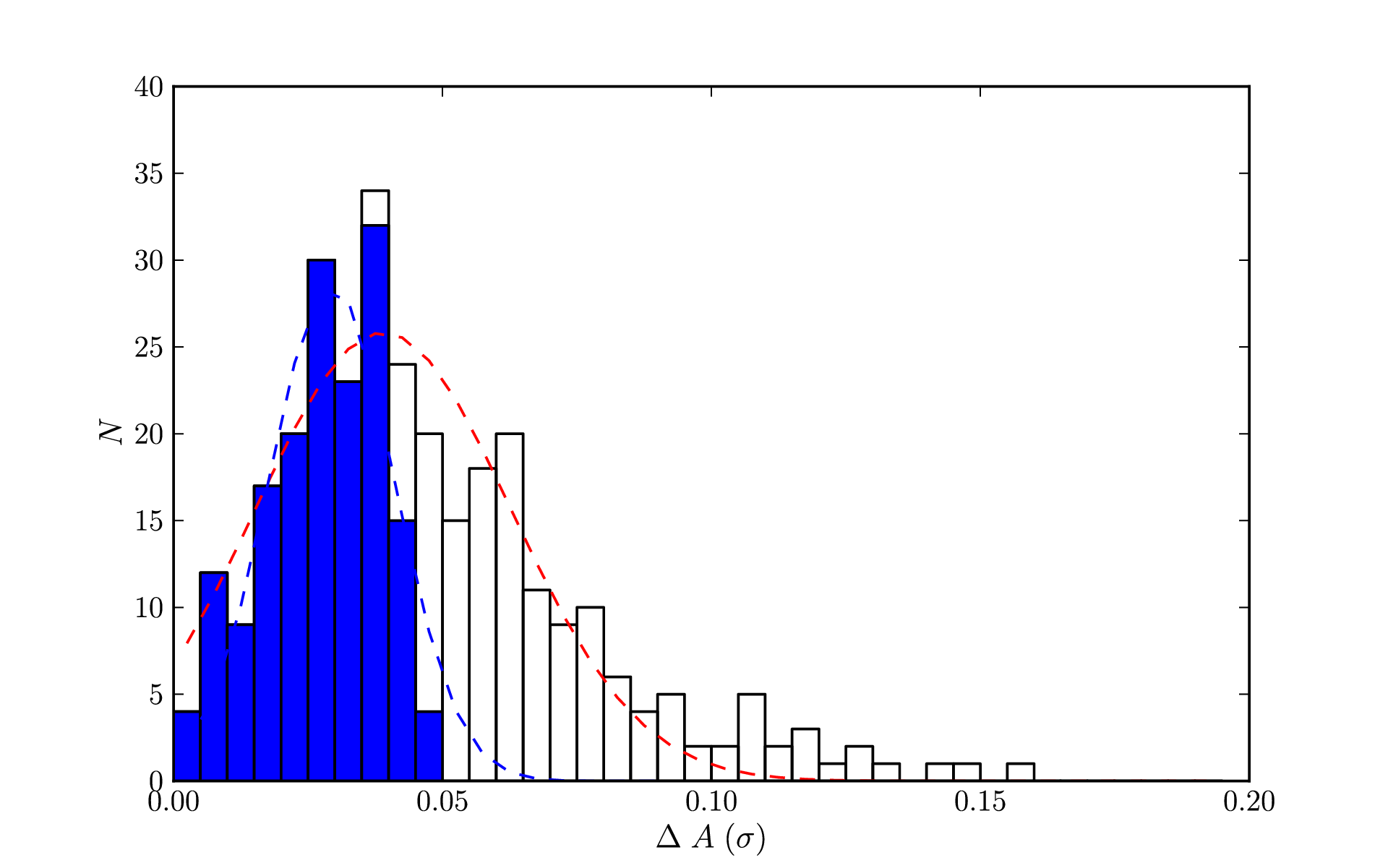}
\includegraphics[width=8.9cm]{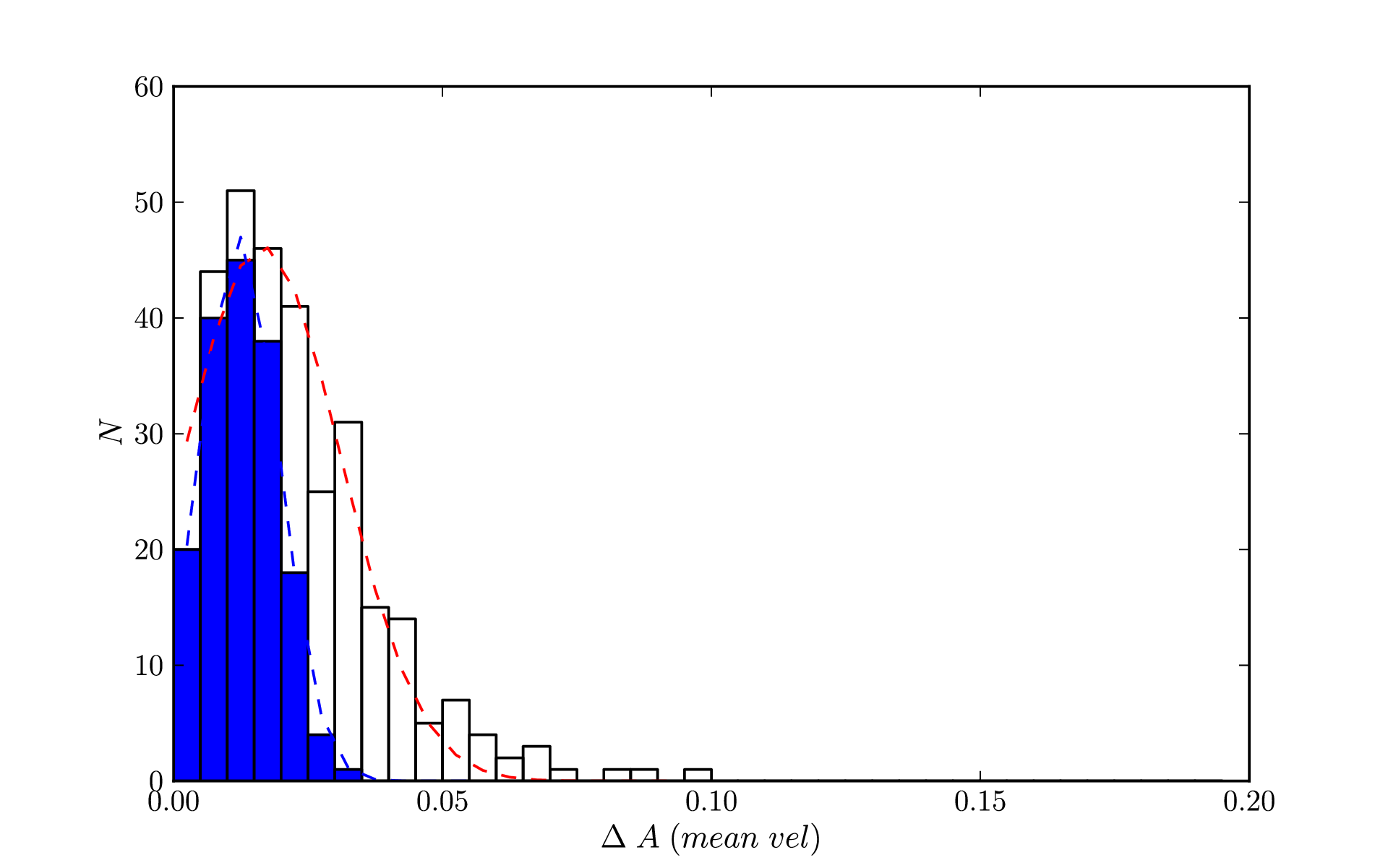}\\
\includegraphics[width=12cm]{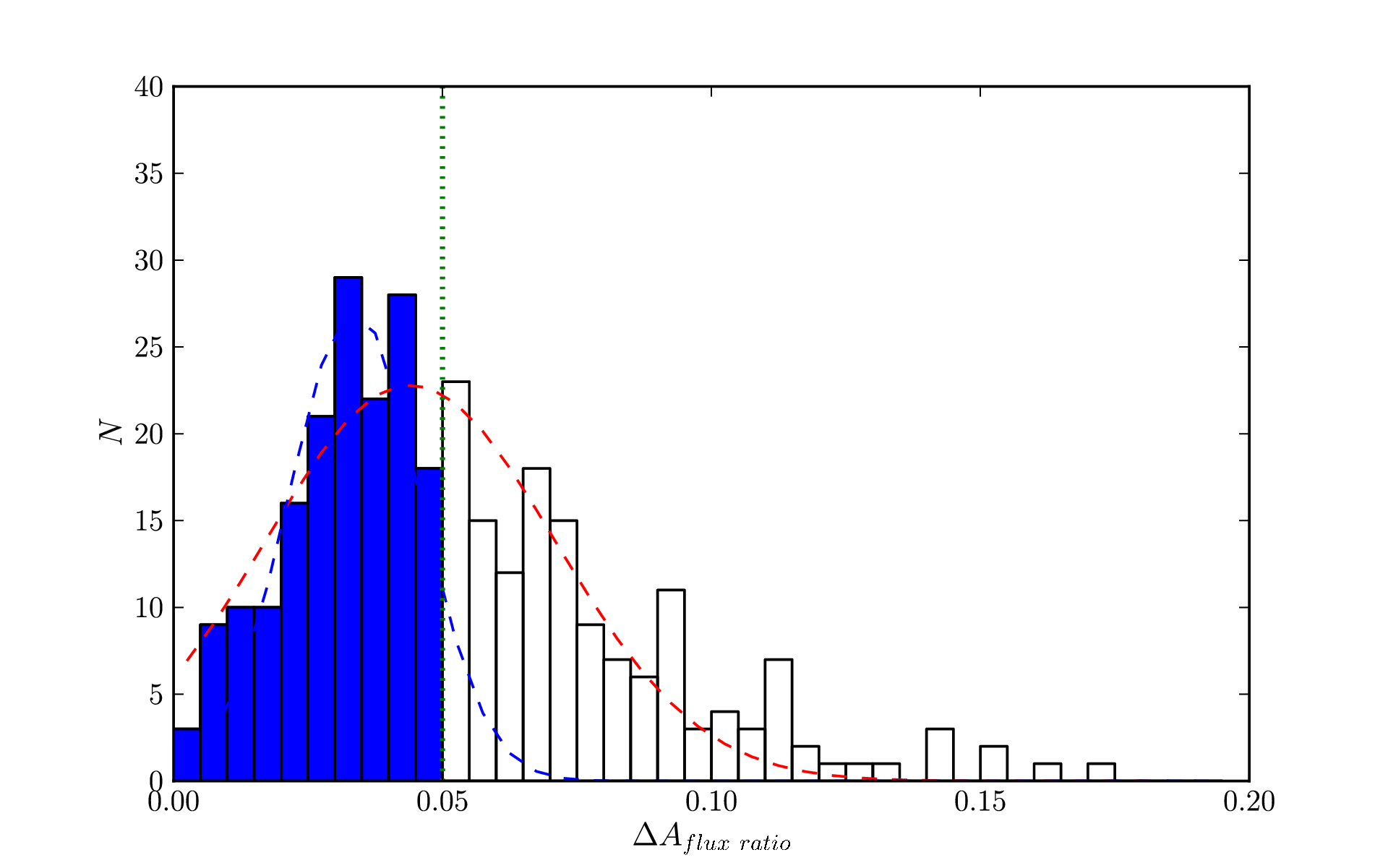}

\caption{ $Upper~right)$ Uncertainty distribution owing to the rms of the \hi\ profiles, $\Delta A(rms)$.  $Upper~left)$  Uncertainty distribution produced by errors in the mean velocity, $\Delta A(mean~vel)$. $Bottom)$ Uncertainty distribution of  $A_{flux~ratio}$ for the \hi\ sample ($N$ = 312) combining the effect of $\Delta A(rms)$, $\Delta A(mean~vel)$ and the small contribution of  $\Delta A(pointing~offset)$ (See Sect.~\ref{sec:meanvel}). 
The \hi\ refined subsample  ($\Delta$$A_{flux~ratio}$ $<$ 0.05, indicated as a dotted line in the lower panel) is shown in the plots as blue filled histograms.
Best Gaussian fits are presented for all distributions as dashed lines: 
for the \hi\ sample:
$a)$ $\mu$=0.02, $\sigma$=0.02,
$b)$ $\mu$=0.04, $\sigma$=0.02, 
and $c)$  $\mu$=0.04, $\sigma$=0.03;
and for the \hi\ refined subsample:
$a)$ $\mu$=0.012, $\sigma$=0.007,
$b)$ $\mu$=0.029, $\sigma$=0.012, 
and $c)$  $\mu$=0.033, $\sigma$=0.012.
\label{fig:uncertainty}}
\end{center}
\end{figure*}

We show in Figure~\ref{fig:uncertainty} $a$ and $b$ the  $\Delta A$\emph{(rms)} and $\Delta A$\emph{(mean vel)} distributions, respectively.  The combined effect of all the previous uncertainties, $ \Delta A_{flux~ratio}$ (including the small contribution of $\Delta A (pointing~offset)$), is shown in Figure~\ref{fig:uncertainty} $c$. We show the best Gaussian fits to the distributions.

\subsection{Comparison between the visual classification and A$_{flux~ratio}$ }
\label{sec:visualVSquant-Afluxratio}

\begin{figure*}
\begin{center}
\includegraphics[width=12cm]{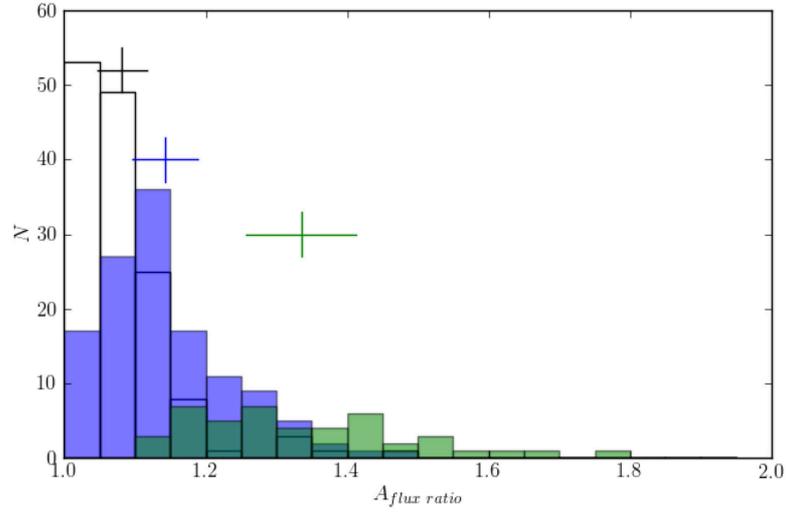}
\caption{Comparison between the visual classification  and the $A_{flux~ratio}$ parameter  for the N = 312 galaxies in the \hi\ sample (symmetric: white histogram, slightly asymmetric: blue or dark gray histogram, and strongly asymmetric: green or light gray histogram). Mean and standard deviations ($\sigma$) for each distribution are also shown by vertical and horizontal lines, respectively. \label{fig:lops-visual-sample-histo-symmetry}}
\end{center}
\end{figure*}

We compare the asymmetry visual classification of the \hi\ profiles (Sect.~\ref{sub:visual})  with the A$_{flux~ratio}$ in
 Figure~\ref{fig:lops-visual-sample-histo-symmetry}.  Three clearly distinct A$_{flux~ratio}$ distributions are seen for those galaxies visually classified as symmetric, slightly asymmetric, and strongly asymmetric (Sect.~\ref{sub:visual}).   The $A_{flux~ratio}$  distribution  of \hi\ profiles  visually classified as symmetric has a mean value equal to 1.08, with a standard deviation of 0.065. The distribution for the slightly asymmetric \hi\ profiles is characterized by a larger mean of 1.13 and a similar standard deviation, 0.09.   The distribution of strongly asymmetric profiles is characterized by a mean of 1.37 and a considerably larger scatter, 0.17, with values as high as $A_{flux~ratio}$ $=$ 1.8. The  $A_{flux~ratio}$ distribution for the slightly asymmetric subsample partially overlaps with those of the symmetric and asymmetric distributions. 
 
 The large overlap  that exhibits the  $A_{flux~ratio}$ distribution for each visually classified subsample is not surprising, because this visual classification is of course subjective, and because the  $A_{flux~ratio}$  parameter misses a few cases where the shape of a real asymmetric profile does not correspond to different areas in the approaching and receding sides. Future work would require to inspect other asymmetry parameters that are sensitive to flag these profiles as asymmetric.

\subsection{\hi\ refined subsample and characterization of the $A_{flux~ratio}$ distribution in a sample of isolated galaxies} 
\label{sec:cleanedsample}

\begin{figure*}
\begin{center}
\includegraphics[width=12cm]{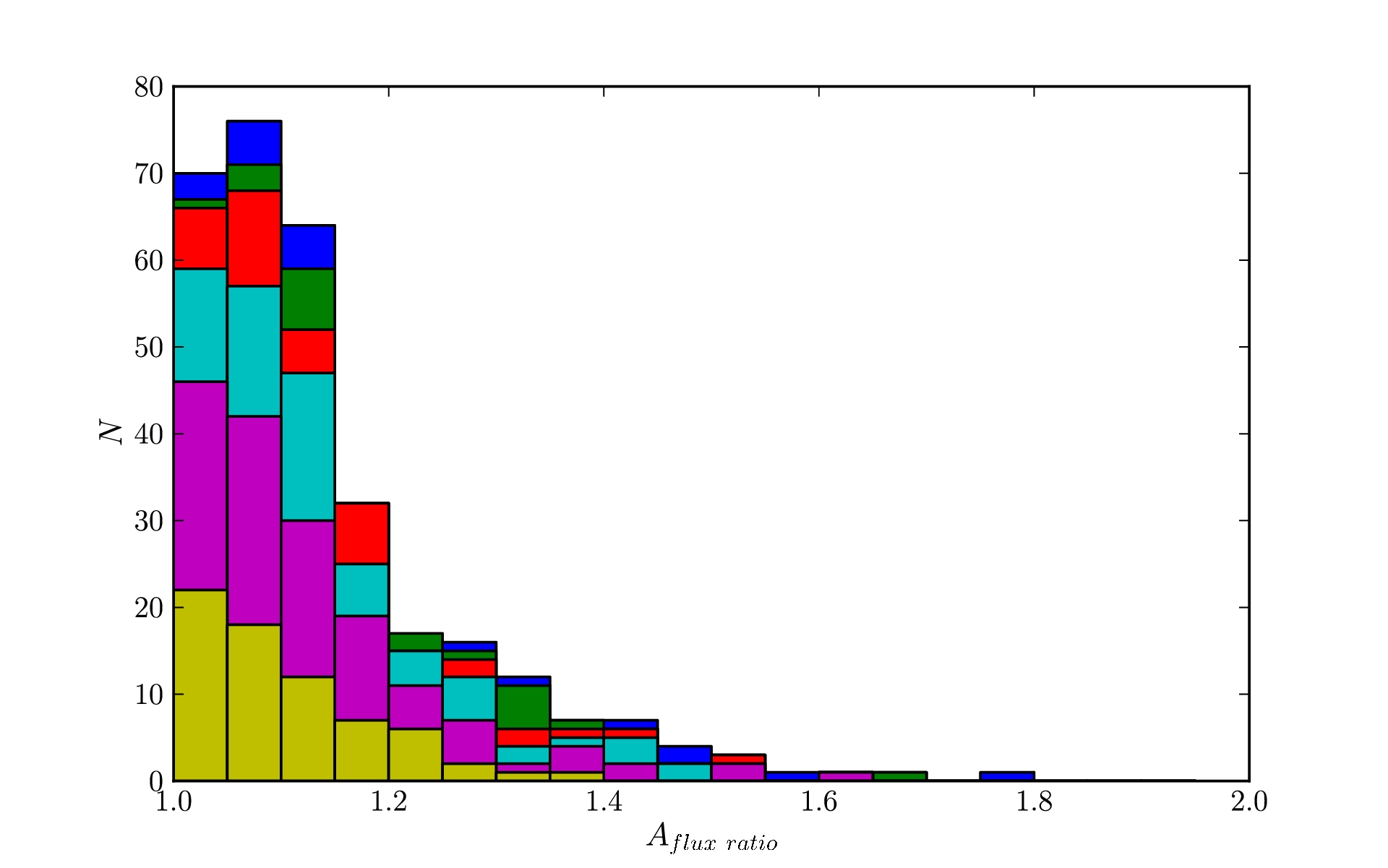}
\caption{
$A_{flux~ratio}$ distribution for the different cuts in $\Delta$$A_{flux~ratio}$ from 0.03 to 0.11  in bins of 0.02. Note that $\Delta$$A_{flux~ratio}$ $<$ 0.05 ($N$ = 166, purple filled histogram) corresponds to the \hi\ refined sample   (see Sect.~\ref{sec:cleanedsample}), and the \hi\ sample ($N$ = 312) is the blue solid line histogram.
\label{fig:cleaning}}
\end{center}
\end{figure*}

The shape of the $A_{flux~ratio}$ distribution might be affected by artificially induced values from the effects explained in Sect.~\ref{sec:meanvel}.  By reducing the net uncertainty in the asymmetry measurement, we reduce errors that might bias our results.
We show in Figure~\ref{fig:cleaning}\ how the $A_{flux~ratio}$ distribution changes for different $\Delta$$A_{flux~ratio}$ limits. The smaller the limit (i.e., only including accurate values of $A_{flux~ratio}$), the better a half-Gaussian reproduces the distribution.

From now on we choose those \hi\ profiles with  $\Delta$$A_{flux~ratio}$ $<$ 0.05, namely the \hi\ refined subsample, to remove from our statistical analysis those profiles with an uncertain determination of the asymmetry index. With this criterion we still have a large sample of $N$ = 166 galaxies. The basic property distributions (velocity, morphological type, $L_B$ and $L_{FIR}$) of the \hi\ refined subsample are shown in Figure~\ref{fig:characterization} as (blue) solid lines, in comparison to those of the \hi\ sample. 

\begin{figure*}
\begin{center}
\includegraphics[width=12cm]{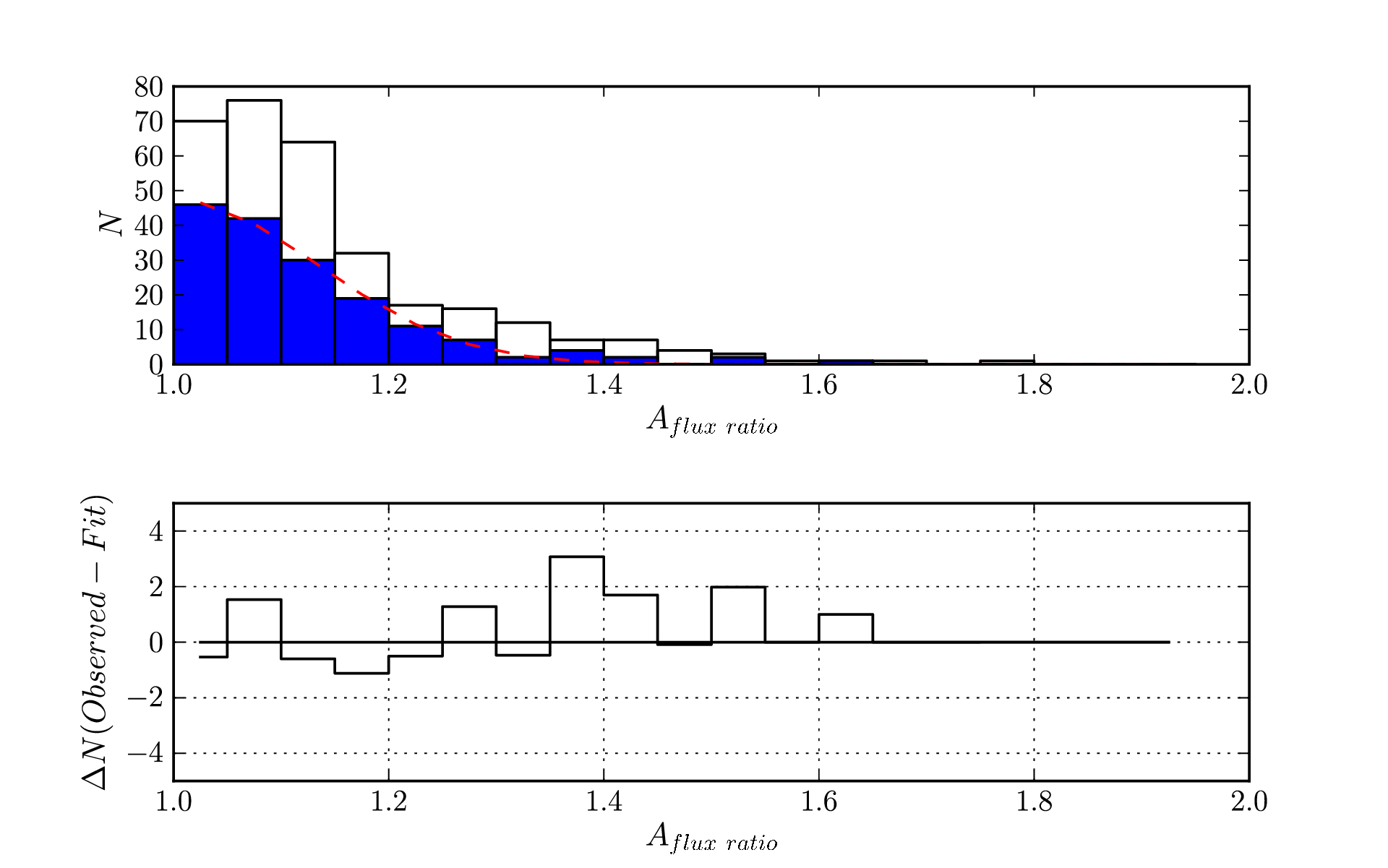}
\caption{ $Upper~panel)$ $A_{flux~ratio}$ distribution of the \hi\ refined subsample  ($\Delta$$A_{flux~ratio}$ $<$ 0.05) ($N$ = 166, blue filled histogram), in comparison with that of the \hi\ sample ($N$ = 312, solid line histogram)   (see Sect.~\ref{sec:cleanedsample}). A half-Gaussian fit (red dashed line) to the \hi\ refined subsample is presented. The half-Gaussian curve is characterized by a standard deviation $\sigma$=0.13, $lower~panel)$ The residual of the half-Gaussian fit to the observed $A_{flux~ratio}$ distribution for the \hi\ refined subsample.
\label{fig:GaussFit}}
\end{center}
\end{figure*}

In order to characterize the intrinsic scatter  of the asymmetry parameter distribution in a sample of
isolated galaxies with minor contamination of artificially asymmetric \hi\ profiles we fitted a half-Gaussian function to the \hi\ refined subsample (Fig~\ref{fig:GaussFit}). The fit yields a width of $\sigma$ = 0.13  (Fig~\ref{fig:GaussFit}). 
This time the fit successfully reproduces the asymmetry parameter distribution, including the low and high ends.
 Only 2\% of the isolated galaxies are in excess of 3$\sigma$.

The width of the half-Gaussian distribution sets an upper limit to the intrinsic dispersion of the \hi\ asymmetry in isolated galaxies. Errors  in the calculation of the asymmetry index might be typically $\sim$ 0.03 (mean of the Gaussian fit) (Sect.~\ref{subsec:presentation}). 
As discussed in Sect.~\ref{subsec:otheruncertainties}, there might also be random errors in the pointing ($\sim$ 0.04) and baseline subtraction ($\sigma$ $\sim$ 0.02) that may increase errors in $A_{flux~ratio}$.  Hence it is reasonable to expect a lower value of the width, $\sigma$ $\sim$ 0.11, once these sources of errors are corrected. 

\begin{figure*}
\includegraphics[width=9.5cm]{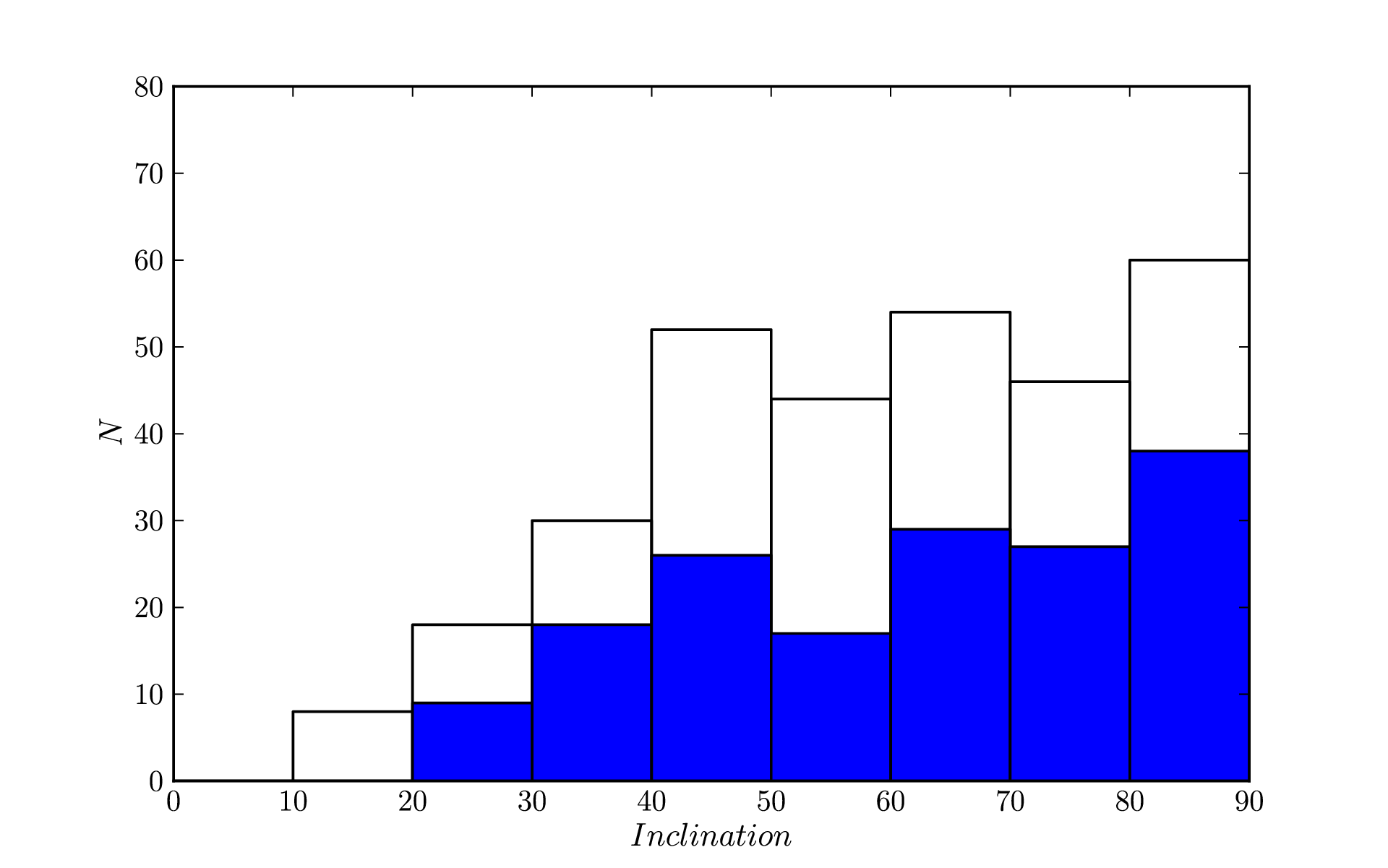}
 \includegraphics[width=9.5cm]{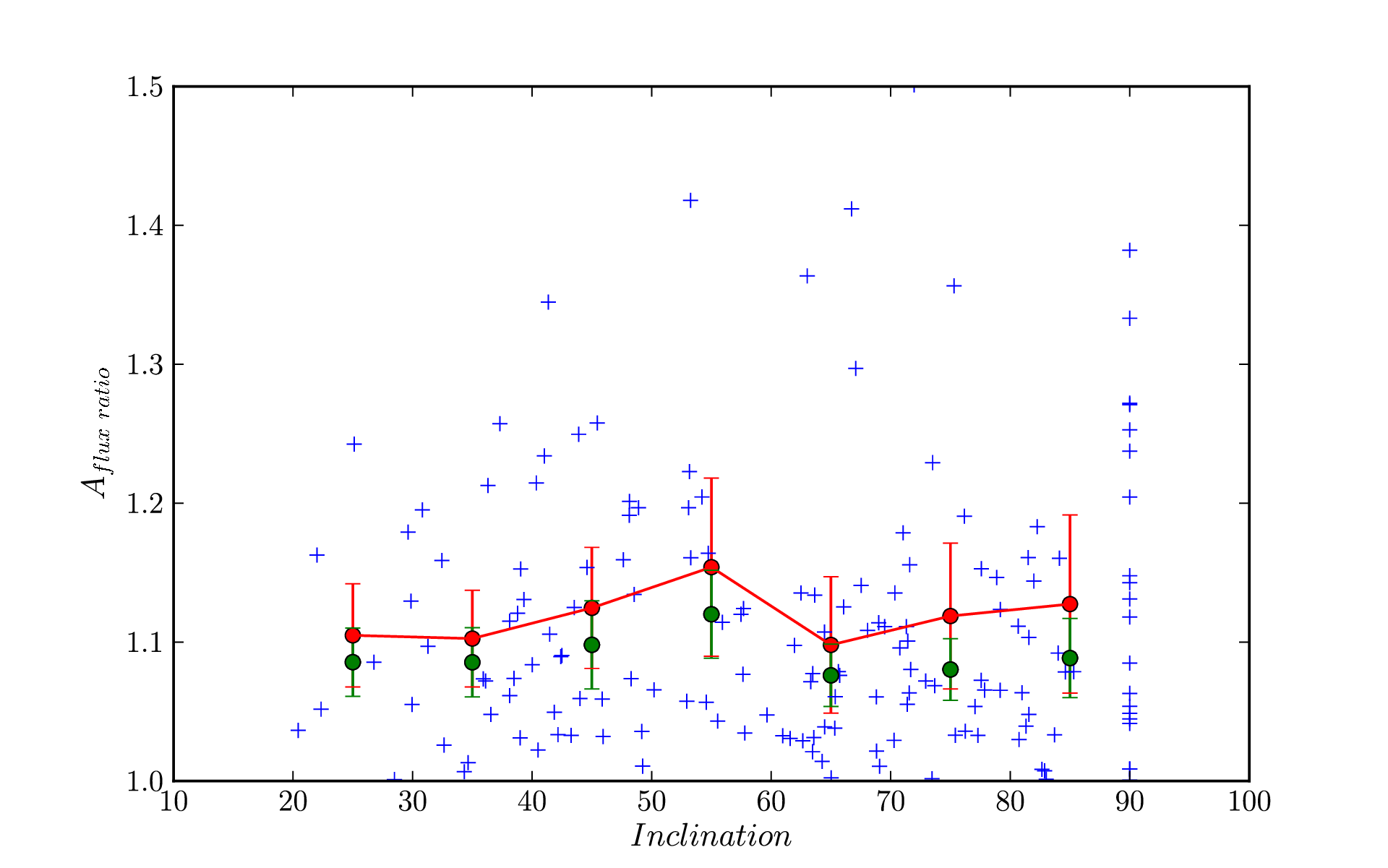}\\
\caption{$a)$ Distribution of inclinations for the N = 312 galaxies in the  \hi\ sample (solid line histogram) and the \hi\ refined subsample (filled histogram). $b)$ $A_{flux~ratio}$ versus  $inclination$, from $i$ = 10 to 90$\arcdeg$ in 10$\arcdeg$ bins for the \hi\ refined subsample. Red points and their error-bars indicate the mean (connected by red solid line) and standard deviation, and green points the median and the median absolute deviation.
\label{fig:inclination}}
\end{figure*}

Note that the quantification of the asymmetry distribution for the galaxies in the \hi\ refined subsample is not affected by inclination effects \citep[e.g.][]{2009PhR...471...75J}.
 Figure~\ref{fig:inclination}$a$ shows that the inclination of the galaxies are distributed homogeneously above $i$ = 30$\rm ^o$.
Only two galaxies have an inclination $i$ $<$ 15$\arcdeg$ (CIG 85 and 178). The lack of galaxies below $i$ = 30 is caused by the width-to-channel ratio criterion explained in Sect.~\ref{sub:sampleselection}.
This homogeneity in the inclination ensures that  most of our galaxies do not show symmetric profiles because the galaxies are viewed face-on, where an asymmetry in the velocity field would remain unnoticed.
To further inspect whether inclination can be introducing any bias in our
results, we plotted the asymmetry index versus the inclination
 (Figure~\ref{fig:inclination}$b$) and found that the two quantities are not correlated.

\section{\hi\ profile lopsidedness and environment}
\label{sec:environment}

\subsection{\hi\ asymmetries and isolation parameters in CIG galaxies}

 A reevaluation and quantification  of isolation degree for CIG galaxies was reported in 
 \citet{2007A&A...470..505V,2007A&A...472..121V}.  \citet{2007A&A...470..505V} derived two 
 isolation parameters for each CIG galaxy: 1) a local surface density parameter $\eta_K$ within 
 the distance to the k-th neighbor (a good tracer of average galaxy surface density) and 2) a
 tidal strength parameter $Q$ (a parameter more sensitive to one-on-one interactions).

\begin{figure*}
\centering
\includegraphics[height=8cm]{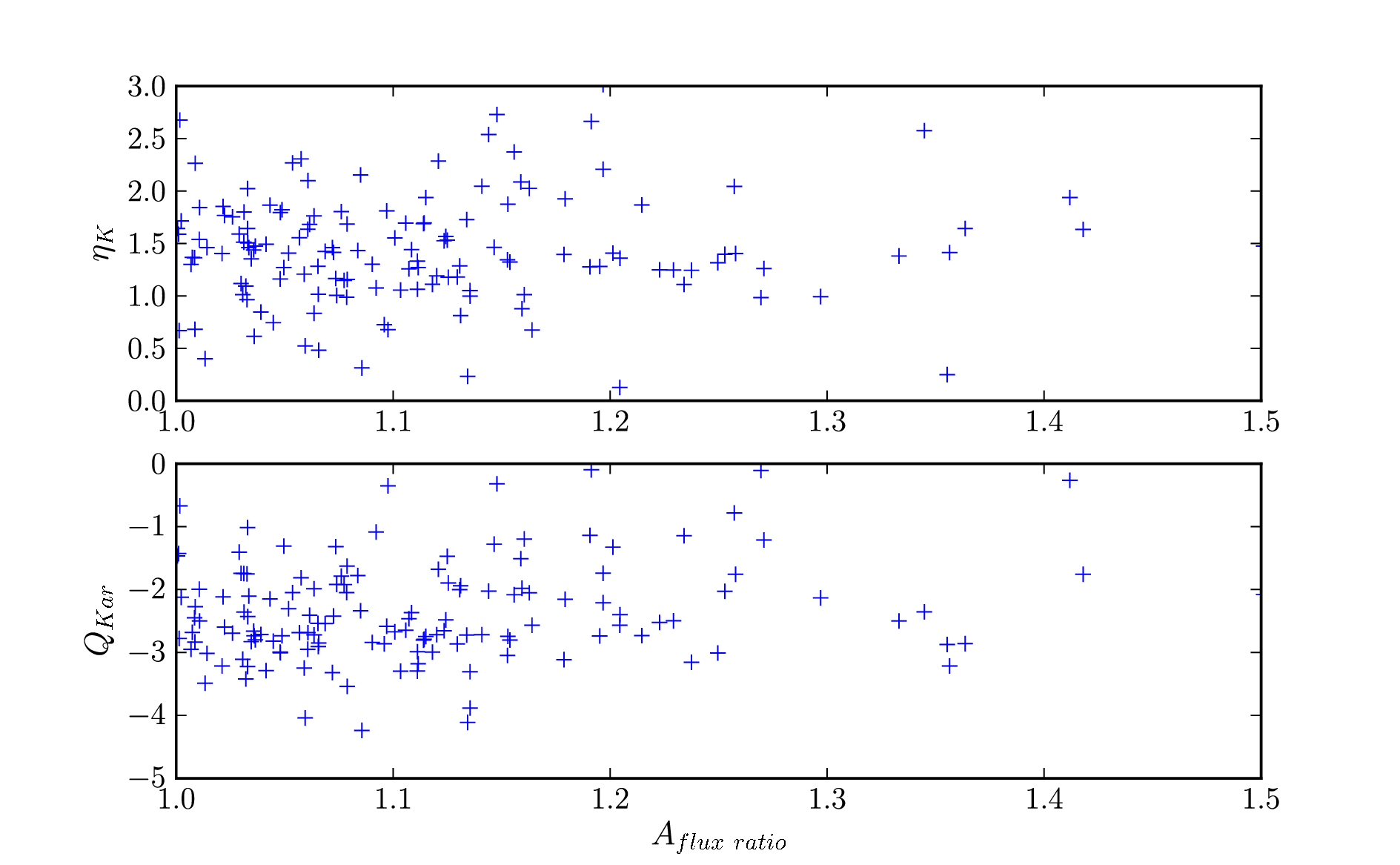}
\caption{
Isolation parameters vs $A_{flux~ratio}$. The isolation parameters \citep{2007A&A...472..121V} are the local number surface density parameter $\eta_K$ to the K-th neighbor, where K = 5 ($upper~panel$), and  the tidal strength parameter $Q$ ($bottom~panel$), which only takes into account similarly size neighbors (factor 4 in size, as defined in  \citealt{1973AISAO...8....3K}).
\label{fig9}}
\end{figure*}

\hi\ is known to be a sensitive diagnostic of interaction motivating us to compare 
these two parameters with our $A_{flux~ratio}$ asymmetry parameter. Figure~\ref{fig9} 
shows the lack of correlation between $A_{flux~ratio}$ and both $\eta_K$ and $Q$.
The Pearson's correlation coefficient is $\rho$ = -0.005 and 0.114, respectively, which 
indicates that the two quantities are essentially not correlated.
A small trend in the $Q$ parameter might be present in the sense that larger \hi\ asymmetries seem to be found 
in less isolated systems. The calculated intercept and slope are -3.3 $\pm$0.7 and 0.8 $\pm$ 0.7, respectively.

The lack of correlation suggest that we are minimizing nurture effects that might affect the \hi\ shape. 
The low values and small range in terms of galaxy density and tidal strength covered by CIG galaxies  
are not enough to see a correlation.

\citet{2005A&A...438..507B} also suggest that there is no correlation between lopsidedness and tidal strength. 
However, they use the lopsidedness $A_1$ parameter on NIR surface density, and NIR emission is not as extended as the \hi . 
 
\subsection{\hi\ asymmetry distribution in field samples}
\label{sub:samples}
We compare the asymmetry distribution of our \hi\ refined subsample with that of different studies from 
the bibliography  where a similar asymmetry index (Sect.~\ref{sec:asymmetry_coefficients}) has been calculated 
and involve field/isolated galaxies (see also Sect.~\ref{sec:introduction}) : \citet{1998AJ....115...62H}, \citet{1998AJ....116.1169M} and \citet{2005A&A...438..507B}. 

\begin{figure*}
\begin{center}
\includegraphics[width=12cm]{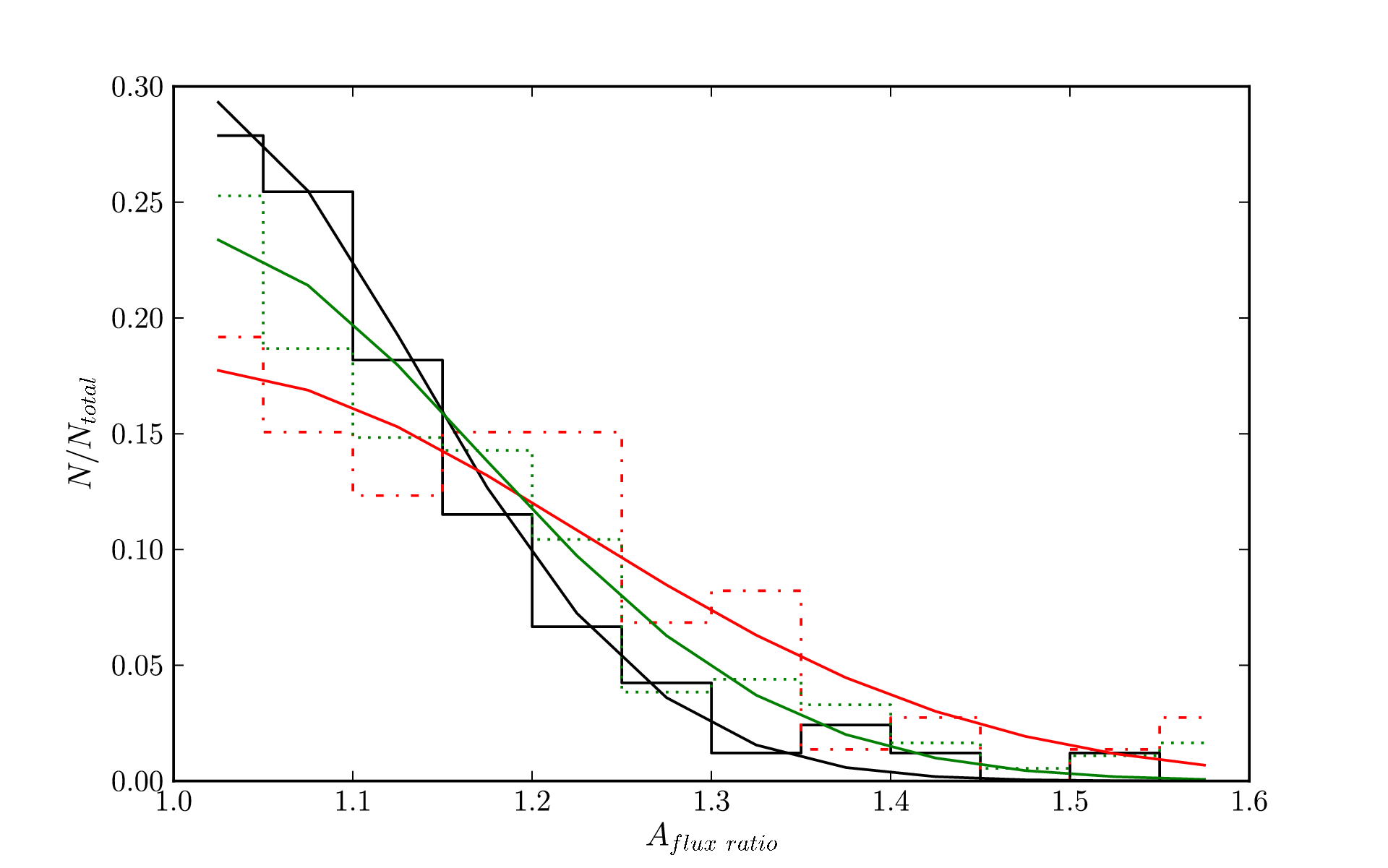}
\end{center}
\caption{Comparison of  the normalized  $A_{flux~ratio}$ distribution between our \hi\ refined sample (black solid histogram) and  1) \citet{2005A&A...438..507B} (red dotted-dashed histogram), 2) a combined sample including \hi\ data in  \citet{2005A&A...438..507B}, \citet{1998AJ....116.1169M} and \citet{1998AJ....115...62H} excluding CIG galaxies (green dotted line). Solid curves are the half-Gaussian curves fitted to each distribution. See Table~\ref{tab:afluxratio} for a comparison of the $\sigma$'s of each half-Gaussian curve. } \label{fig:comparison_hg98-2}
\end{figure*}

Figure~\ref{fig:comparison_hg98-2} shows the $A_{flux~ratio}$ normalized distribution for our refined subsample with 1)  \citet{2005A&A...438..507B} and  2) a combined sample (N = 186) including \hi\ data in \citet{1998AJ....116.1169M}, \citet{2005A&A...438..507B}, and \citet{1998AJ....115...62H} excluding CIG galaxies (80 galaxies).  
The \hi\ refined subsample shows the distribution best described by a half-Gaussian. It also shows the lowest absolute value of $\sigma$. 
The \citet{2005A&A...438..507B} distribution shows the widest distribution ($\sigma$ = 0.23) and noticeably deviates from 
a half-Gaussian curve. An intermediate case, $\sigma$ = 0.17, is found for the combined sample without CIG galaxies.
Table~\ref{tab:afluxratio} gives $\sigma$ values for each distribution as well as an asymmetry rate with ''asymmetric''
profiles defined as $A_{flux~ratio}$ values exceeding the 2$\sigma$ level of our \hi\ refined subsample ($A_{flux~ratio}$ = 1.26).

\begin{figure*}
\begin{center}
\includegraphics[width=16cm]{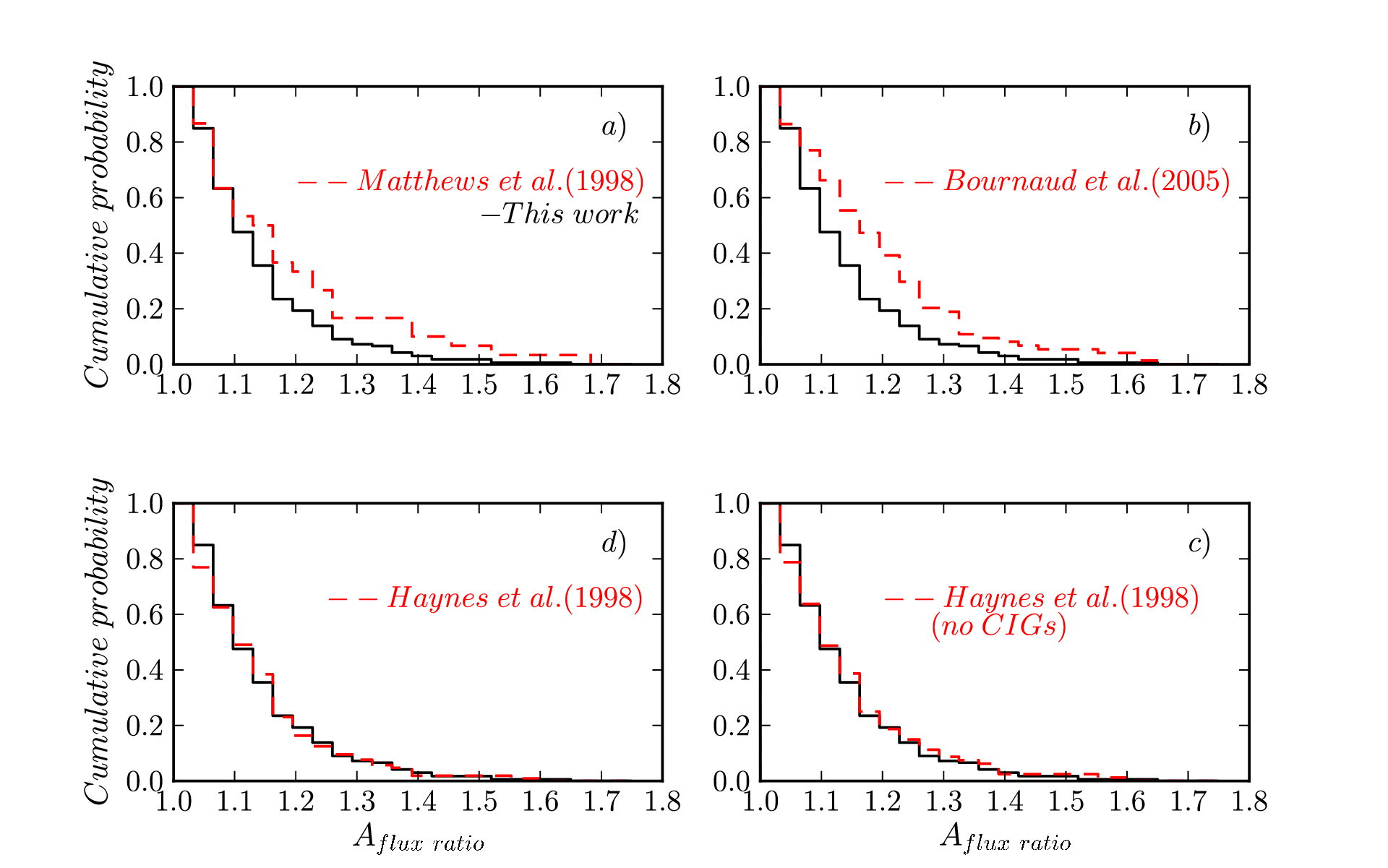}
\end{center}
\caption{Comparison of  the cumulative  $A_{flux~ratio}$ distribution between our sample (black solid line) and other samples (red dashed lines): $a)$ \citet[][$N$ = 30]{1998AJ....116.1169M}, $b)$ \citet[][$N$ = 76]{2005A&A...438..507B}, $c)$  \citet[][$N$ = 104)]{1998AJ....115...62H} ($N$ = 106), and $d)$ \citet{1998AJ....115...62H} excluding CIG galaxies ($N$ = 80). See Table~\ref{tab:afluxratio} for an asymmetry rate comparison at a $A_{flux~ratio}$ = 1.26 level.} \label{fig:comparison_hg98}
\end{figure*}

Figure~\ref{fig:comparison_hg98} compares the $A_{flux~ratio}$ cumulative probability distribution for our \hi\ refined sample and those of \citet{1998AJ....115...62H}, \citet{1998AJ....116.1169M}, and \citet{2005A&A...438..507B}. 
In each plot the difference of the two curves indicates the asymmetry rate difference for a given $A_{flux~ratio}$ limit.
Our sample lies below the field samples in almost every bin with differences typically between 10 -- 20\%.
A result more similar to our sample is found for  \citet{1998AJ....115...62H} likely in part because of the significant
fraction of CIG galaxies (23\%) included in their sample. Removing the CIG overlap increases both their 
$\sigma$ and asymmetry rate.

 We performed a $\chi^2$ test to check whether the null hypothesis that any of the three $A_{flux~ratio}$ distributions 
 is similar to our \hi\ refined sample, could be rejected. Except for \citet{1998AJ....115...62H} ($\chi^2$ = 9 and the 
 associated p-value = 0.33) this hypothesis can be rejected. In the cases of \citet{2005A&A...438..507B} and 
 \citet{1998AJ....116.1169M}, $\chi^2$ $=$ 47 (p-value $=$ 2 $\times$ 10$^{-7}$) and $\chi^2$ $=$ 14 (p-value $=$ 0.09) 
 respectively. The sample differences we find are significant and cannot be ascribed to the refinement of the \hi\ sample (Sect.~\ref{sec:cleanedsample}). In principle we do not know how much  \citet{1998AJ....116.1169M} and \citet{2005A&A...438..507B}'s observations are affected by systematic errors, but  differences involving the same criterion as used in our study would yield an asymmetry rate difference $<$ 5\%. \citet{1998AJ....115...62H} included only high S/N profiles, avoided pointing problems and quantified baseline problems, suggesting it is reasonable to compare it directly with our \hi\ refined sample.
Overall, because of their degree of isolation, our \hi\ refined subsample and \citeauthor{1998AJ....115...62H}'s  show a lower 
frequency ($\sim$ 10 -- 20\%) of galaxies with asymmetric profiles than in other samples such as \citet{1998AJ....116.1169M} and \citet{2005A&A...438..507B}.

 \section{Relation between \hi\ profile lopsidedness and optical/FIR properties}
 \label{sec:optical}

 In this section we  explore possible correlations between the asymmetry index $A_{flux~ratio}$  and optical properties 
 of the \hi\ refined subsample such as morphology, optical signs of perturbation, optical luminosity ($L_B$), and far-infrared luminosity ($L_{FIR}$).

   \subsection{Morphology and luminosity}
\label{sec:morphology}

\begin{figure*}
 \begin{center}
 \includegraphics[width=12cm]{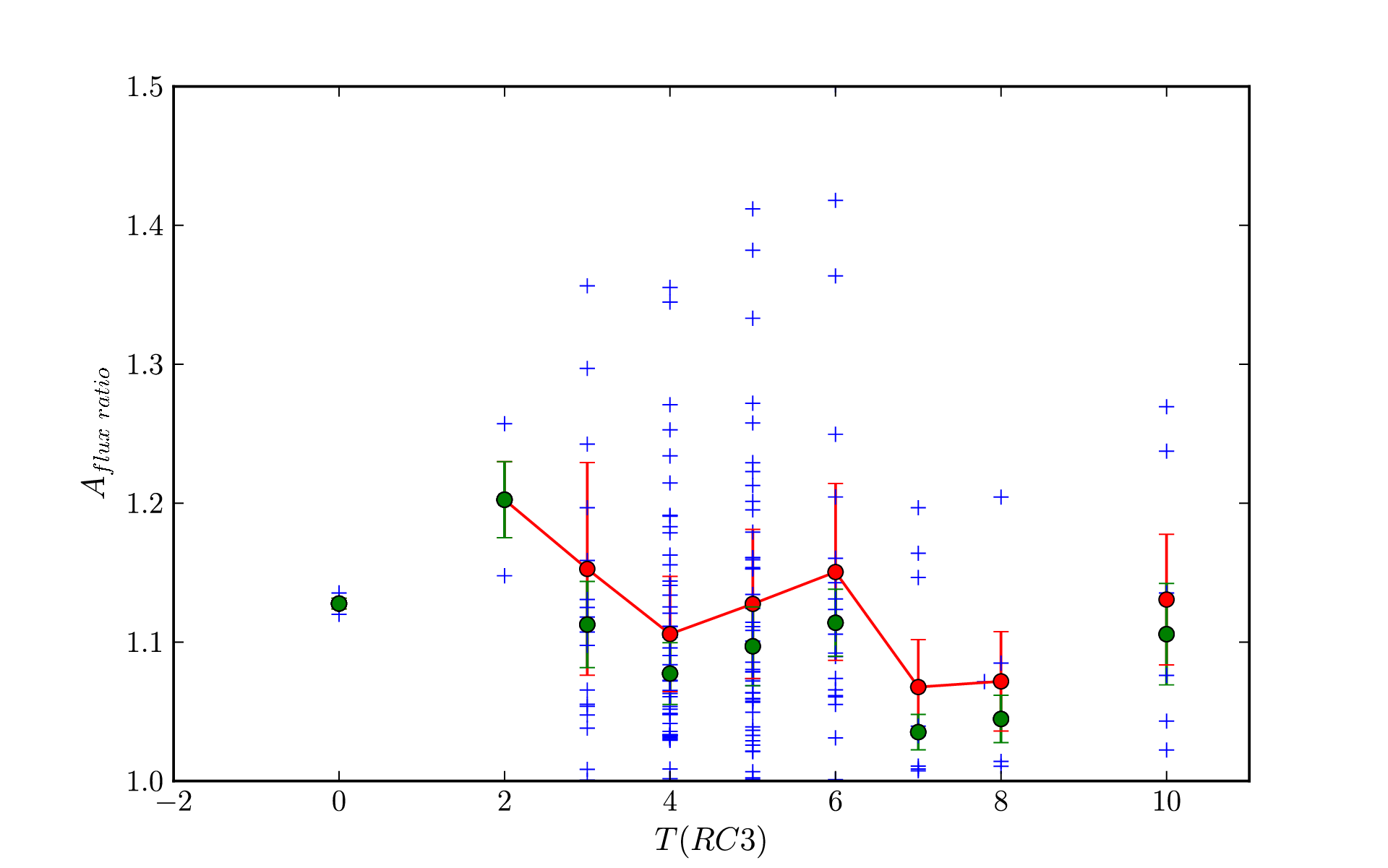}
 \end{center}
 \caption{ $A_{flux~ratio}$ and  $T(RC3)$  \citep{2006A&A...449..937S}  for the \hi\ refined subsample. Red points and their error-bars indicate the mean (connected by red solid line) and standard deviation, and green points the median and the median absolute deviation, for each morphological type from $T(RC3)$ = -5 to 10 (E to Im).  \label{fig:afluxratiomorphologya}}
 \end{figure*}

Figure~\ref{fig:afluxratiomorphologya} shows the  distribution of $A_{flux~ratio}$ values for each Hubble morphological 
class (median, mean, and standard deviation values are indicated). The bins representing the majority of our sample 
($T(RC3)$ = 3 to 6, i.e. Sb to Scd) show a fairly large scatter (standard deviation $\sigma$$\sim$ 0.1). We see a slight decreasing 
trend in $A_{flux~ratio}$ toward later-type galaxies. 

Studies of the relation between \hi\ lopsidedness and morphological type are rare.
\citet{1998AJ....116.1169M} studied a sample of 30 moderate to low surface brightness (T=6--9) galaxies
and found a higher asymmetry rate than for more luminous (and higher surface brightness) late-type 
spirals. They found (for this type range) that later types were more likely to show larger asymmetries. In the Eridanus group the $A_1$ parameter as calculated from \hi\ maps  is larger for earlier type galaxies, suggesting that tidal interactions generate a higher lopsidedness rate in galaxies undergoing secular evolution toward earlier type \citep{2006MNRAS.369.1849A}. This result (i.e., larger asymmetries for earlier types) agrees with the general trend seen in our sample.

\begin{figure*}
 \begin{center}
  \includegraphics[width=8.5cm]{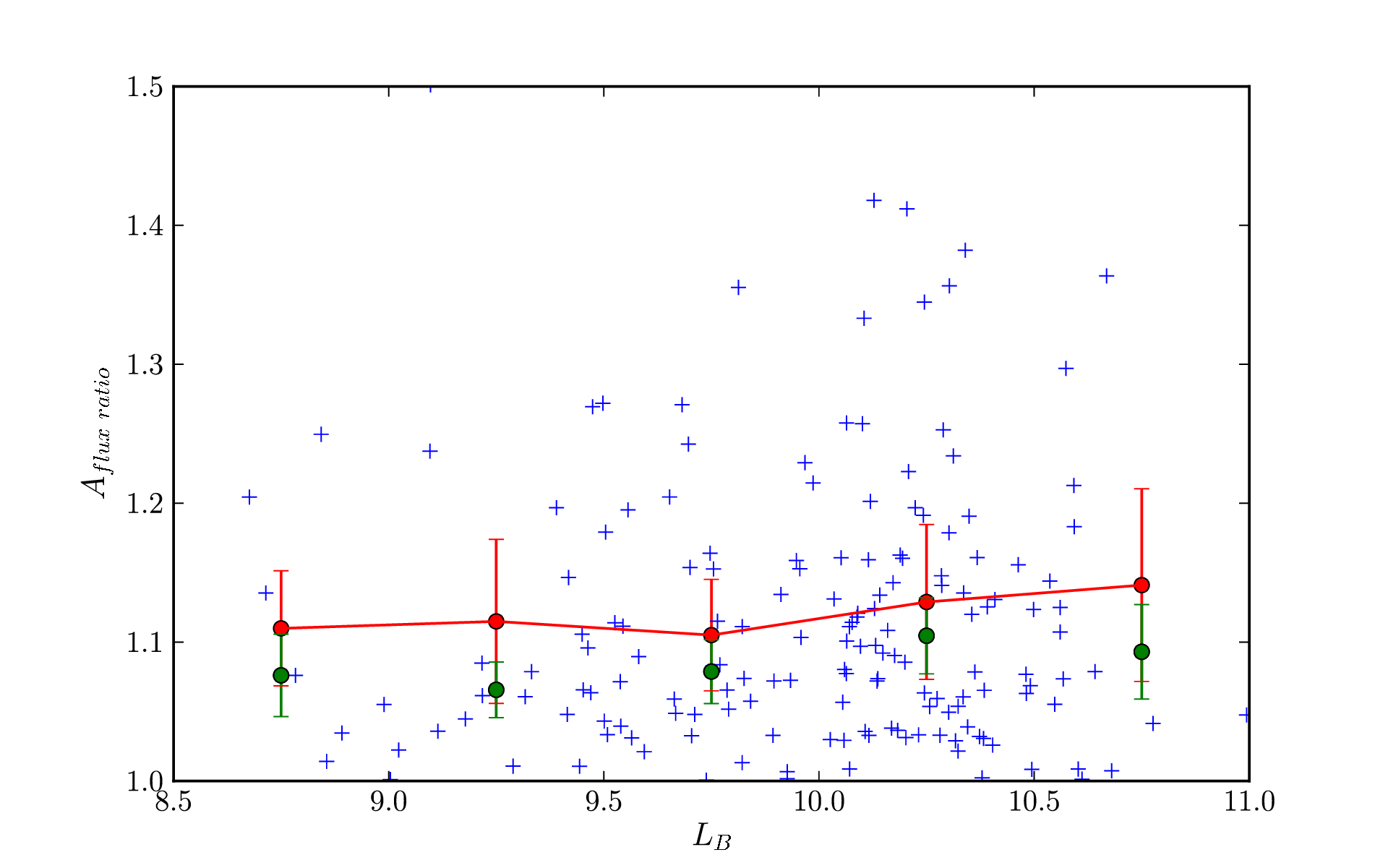} 
 \includegraphics[width=8.5cm]{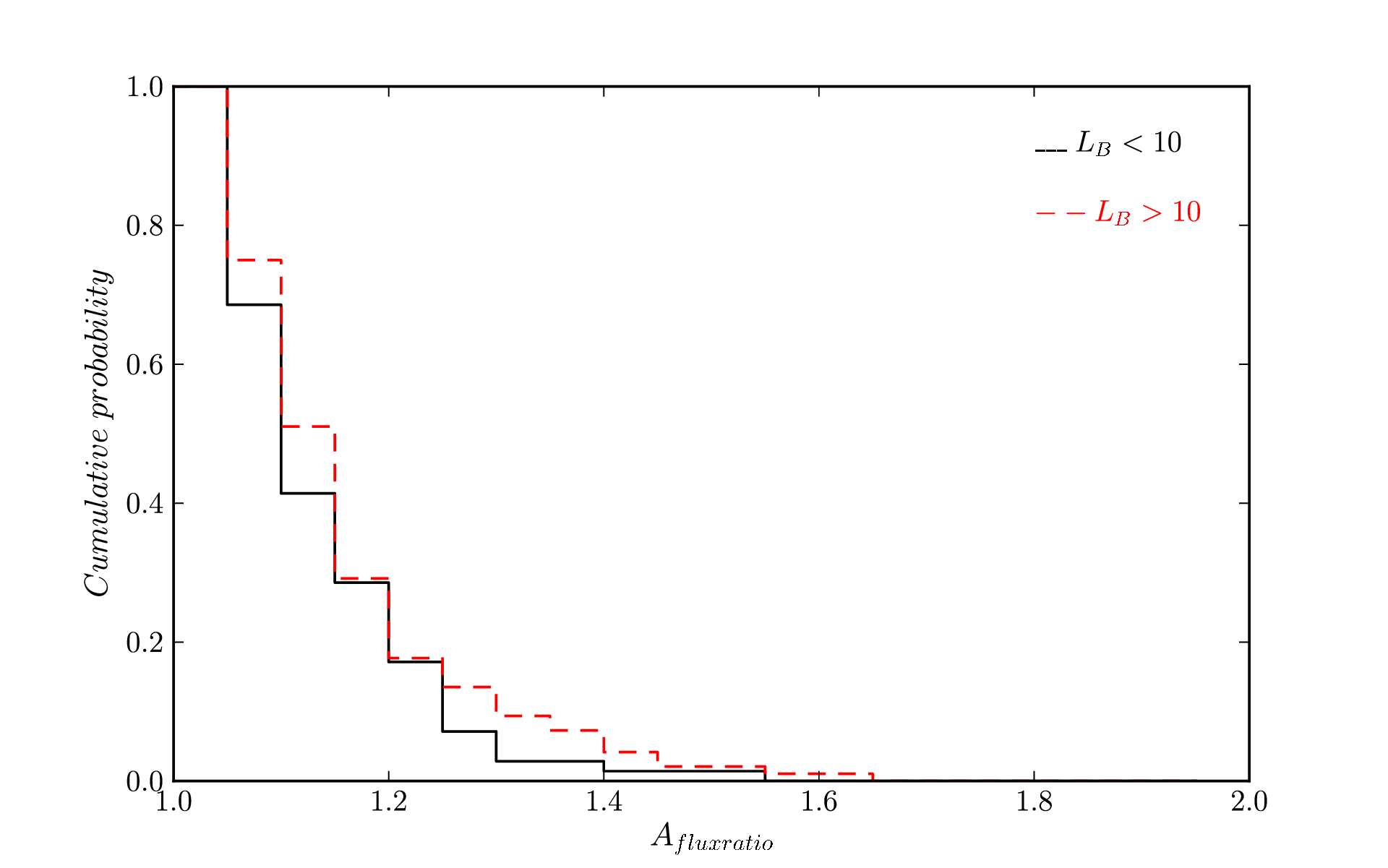}
 \end{center}
 \caption{ $Left)$ $A_{flux~ratio}$ versus log($L_B$[$L_\odot$]). Symbols are as in  Figure~\ref{fig:afluxratiomorphologya}. $Right)$ Cumulative probability distribution of  $A_{flux~ratio}$ for log($L_B$[$L_\odot$]) $<$ 10  (solid line) and  log($L_B$[$L_\odot$]) $>$ 10  (dashed red line), using the \hi\ refined subsample. 
  \label{fig:afluxratiomorphologyb}}
 \end{figure*}

We compared the $A_{flux~ratio}$ with luminosity.
More luminous galaxies are slightly more asymmetric (Figure~\ref{fig:afluxratiomorphologyb}, left panel).
Figure~\ref{fig:afluxratiomorphologyb} (right panel) presents the cumulative distribution of $A_{flux~ratio}$ for the 
high- and low-luminosity subsamples \citep{2005A&A...436..443V}. 
The two distributions are different at a level of $\alpha$ $=$ 0.05 using a chi square test: $\chi^2$=14 and $p-value$=0.01. 

  \subsection{Optical signs of interactions}
\label{sec:opticalinteracions}

Here we inspect a possible connection between optical signs of interaction and  asymmetries in the \hi\ profiles.
 \citet{2006A&A...449..937S} revised  the optical morphology classification for the CIG sample using POSS2/SDSS data. Although the CIG, the starting sample of AMIGA, has been selected to minimize close neighbors to the target galaxy and thus interactions, still  \citeauthor{2006A&A...449..937S}'s revision revealed $N$ = 193 objects with nearby companions or signs of distortion likely caused by an interaction.
\citet{2006A&A...449..937S} flagged these galaxies as $interacting$ in the case of a morphologically distorted system and/or almost certain interacting system or flagged as $possibly~interacting$ if there was any evidence of interaction/asymmetry with/without certain detection of a close companion. 

There is no statistically significant difference in terms of \hi\ profile asymmetry rate between galaxies that are optically classified as interacting and those without any sign of interaction. This result is consistent with the conclusion by \citet{2000AJ....120..139K} and \citet{2004AJ....127.1900W} that optical asymmetries are not necessarily correlated with a lopsided \hi\ component.

\subsection{Bar and spiral strengths}

\begin{figure*}
\centering
\includegraphics[width=12cm]{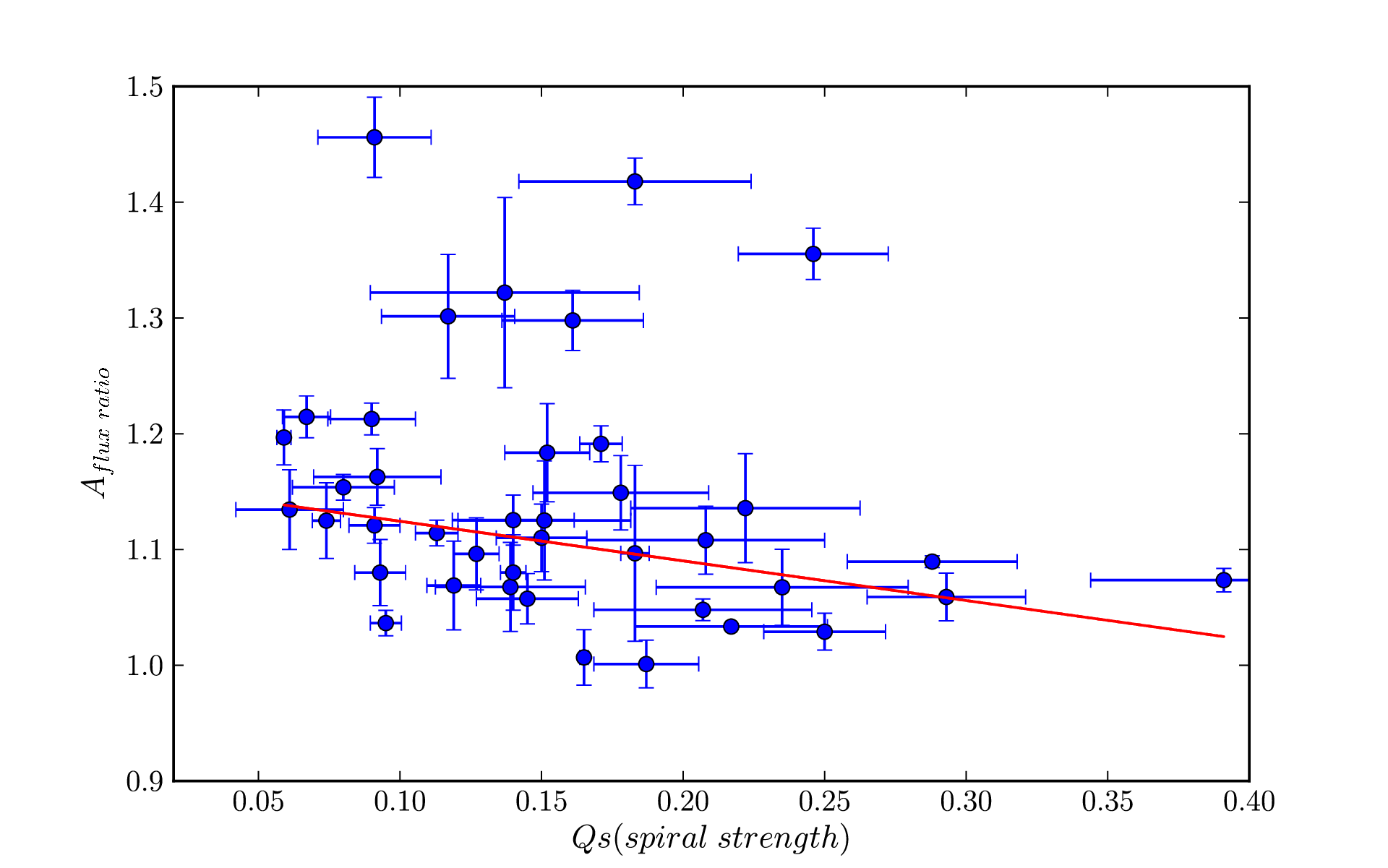}
\caption{$A_{flux~ratio}$ versus spiral  strength ($Qs$) for 40 CIG galaxies  overlapping between the \hi\ sample and the CIG galaxies in \citet{2009AAS...21344310D}. The fit to the data points (slope and intercept are -0.34 and 1.16, Pearson's correlation coefficient $\rho$ = -0.45) is shown as a (red) solid line. The six outliers have been ignored in this fit.}
\label{fig:strengths}
\end{figure*}

We compared the \hi\ asymmetry parameter with the relative spiral and bar strengths calculated  as the maximal torque, or ratio of the maximum tangential force and the azimuthally averaged radial force. This has been obtained for a subsample of N = 96 CIG galaxies by \citet{2009AAS...21344310D} using Fourier analysis over spiral and bar components separately \citep{2003AJ....126.1148B}. Figure~\ref{fig:strengths} shows the $A_{flux~ratio}$ parameter with respect to the relative spiral strength ($Q_s$). The overlapping sample between our \hi\ refined subsample and the one used by  \citet{2009AAS...21344310D} is composed of 40 galaxies.
$Q_s$ seems to anti-correlate with $A_{flux~ratio}$: disks with weaker spiral arms show stronger asymmetries 
(Figure~\ref{fig:strengths}). There are six galaxies (CIGs 11, 33, 689, 712, 912, and 931) that are outliers to this relation in the high end of the $A_{flux~ratio}$ parameter. These are likely galaxies whose \hi\ asymmetry parameter is affected by instrumental effects.
 On the other hand we do not find any correlation of $A_{flux~ratio}$ and $Q_b$.
The total strength $Q_g$ is not correlated with $A_{flux~ratio}$ either, not surprisingly because $Q_b$ and $Q_g$ presents a good correlation \citep{2009AAS...21344310D}.
 The relation between $Q_s$ and $A_{flux~ratio}$ may originate in the observed trends in $Q_s$ versus $T(RC3)$, because $Q_s$ would be expected to correlate with the latter. However, we did not find a clear trend between $Q_s$ and $T(RC3)$. 

The correlation found between $A_{flux~ratio}$ and  $Q_s$ suggests that other samples of galaxies characterized by lower spiral strengths  will have higher \hi\ asymmetry rates. This is consistent with the OSU sample \citep{2002ApJS..143...73E}, where spiral strength is lower than for the CIG (median equal to 0.132 versus 0.161, respectively)  \citep{2005AJ....130..506B,2009AAS...21344310D}, and have a higher \hi\ asymmetry rate with respect to the \hi\ refined subsample  \citep[see Sect.~\ref{sub:samples}, and ][]{2005A&A...438..507B}.

\subsection{FIR luminosities ($L_{FIR}$)}
\begin{figure*}
 \begin{center}
  \includegraphics[width=8.5cm]{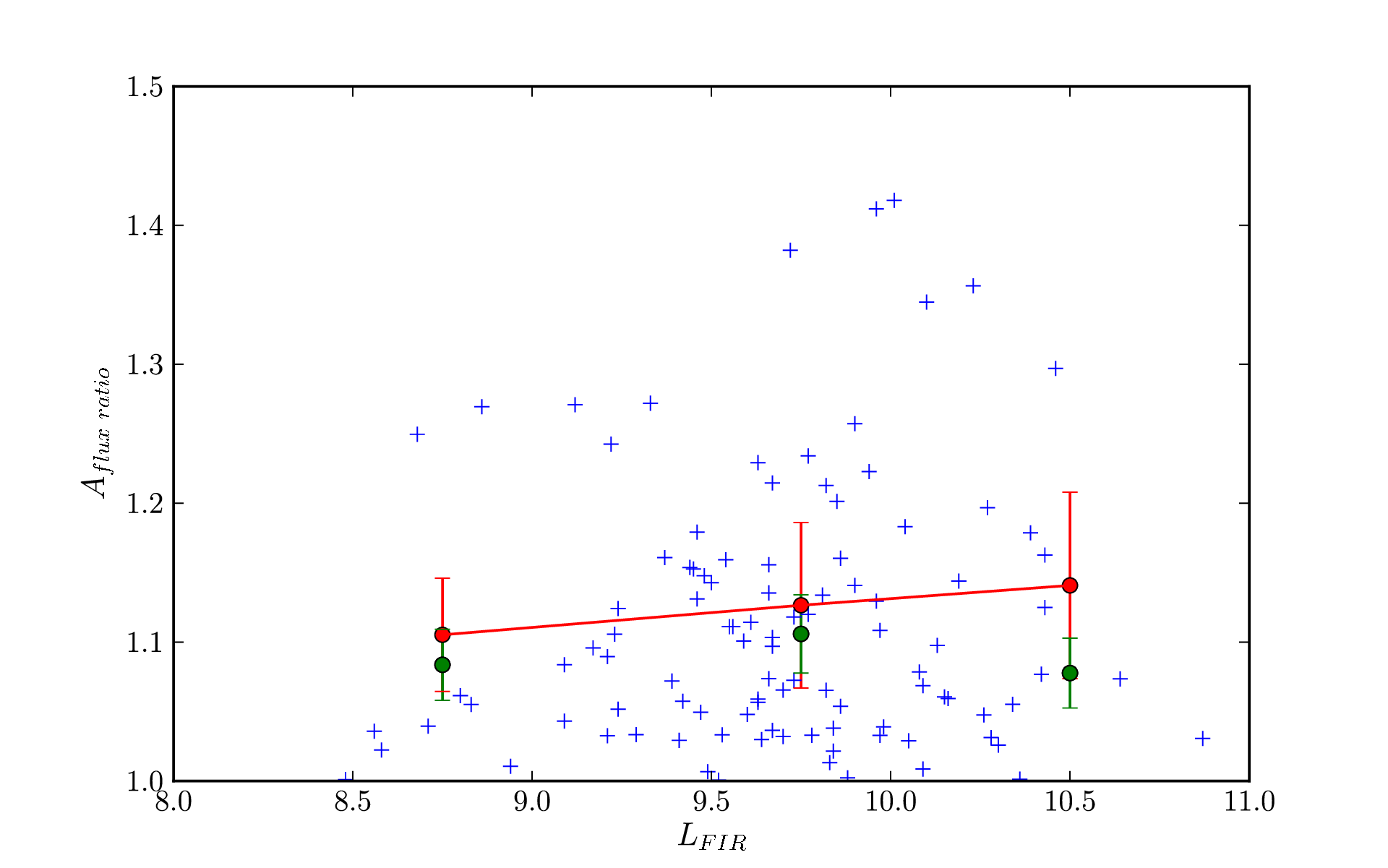}
 \includegraphics[width=8.5cm]{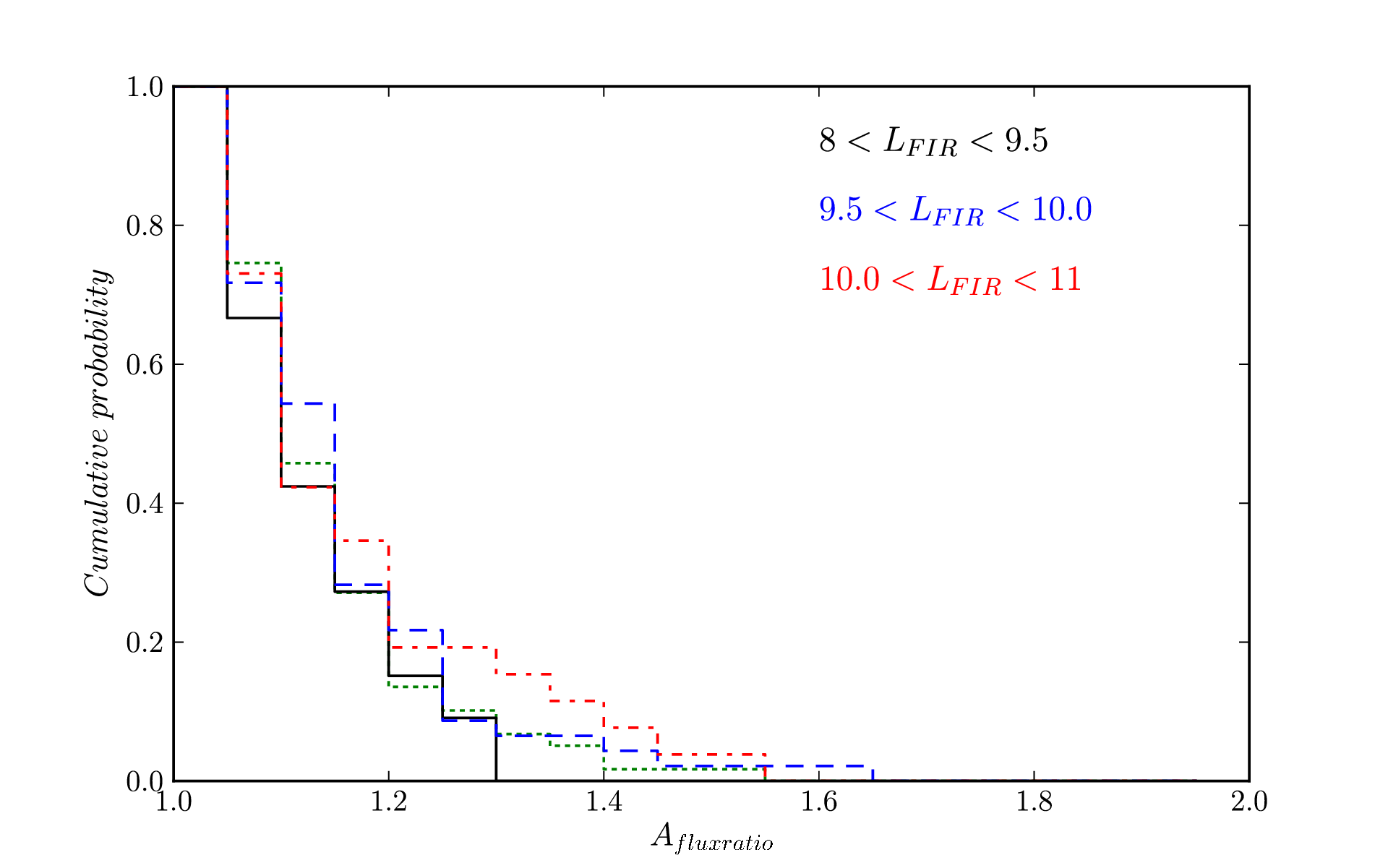}
 \end{center}
 \caption{$Left)$ $A_{flux~ratio}$ versus log($L_{FIR}$[$L_\odot$]). Symbols are as in  Figure~\ref{fig:afluxratiomorphologya}. $Right)$ $A_{flux~ratio}$ cumulative probability distribution for three $L_{FIR}$ bins:  9 $<$ log($L_{FIR}$[$L_\odot$]) $<$ 9.5 (solid line),  9.5 $<$ log($L_{FIR}$ [$L_\odot$]) $<$ 10.0 (blue dashed line) and   10.0 $<$ log($L_{FIR}$[$L_\odot$]) $<$ 11.0  (dash-dotted line). The (green) dotted line represents the $A_{flux~ratio}$ cumulative distribution for those galaxies with an upper limit. \label{fig:afluxratiomorphologyc}}.
 \end{figure*}

$L_{FIR}$ is a good tracer of the star-formation rate and is related to the environment 
in the sense that IR luminous galaxies ($L_{FIR}$ $>$ 10$^{11}$ L$_\odot$) are usually 
interacting or merger systems \citep{1996ARA&A..34..749S}. Unlike other samples of galaxies, 
our isolated population shows low FIR measures e.g. log($L_{FIR}$) peaks from 9.0 -- 10.5 
with very few ($<$2\%) galaxies above 10.5 \citep{2007A&A...462..507L}. The low $L_{FIR}$ values 
of the CIG sample support our claim that the revised CIG (AMIGA) is a sample with only
isolated systems \citep{2007A&A...462..507L}. Here we inspect whether the small fraction of 
IR-luminous systems in our sample corresponds to galaxies with larger \hi\ asymmetries.
We find $N$=165 galaxies in our refined \hi\ sample with IRAS measures, including upper limits. 
Figure~\ref{fig:afluxratiomorphologyc} (left panel) presents a slight trend in the sense that more luminous FIR galaxies have more asymmetric \hi\ profiles.
Right panel of Figure~\ref{fig:afluxratiomorphologyc} shows the  cumulative probability distribution for $A_{flux~ratio}$
in three bins: 8.0~$<$~log($L_{FIR}$[$L_\odot$])~$<$~9.5, 9.5~$<$~log($L_{FIR}$[$L_\odot$])~$<$~10.0 and 10.0~$<$~log($L_{FIR}$[$L_\odot$])~$<$~11.0.  We also show the distribution for those 
galaxies  ($N$= 60) with FIR upper limits. The null-hypothesis that the first two bins  are similar cannot be rejected (at a level of $\alpha$ $=$ 0.05) using a $\chi^2$ square test: $\chi^2$=6 and $p-value$=0.24. 
On the other hand we find that the  $A_{flux~ratio}$ distributions for the latter two bins are different 
($\chi^2$=14, $p-value$=0.008) from each other. We find a 10-20\% 
excess of higher asymmetry values for the most FIR luminous galaxies (10.0~$<$~log($L_{FIR}$[$L_\odot$])~$<$~11.0).  
If real, this excess might reflect asymmetries and FIR luminosities simultaneously enhanced by accretion events.

\section{Discussion}
\label{sec:discussion}

The CIG sample represents the $\sim$3\% most isolated galaxies in the Catalog of Galaxies and Clusters of Galaxies (CGCG, \citealt{1961cgcg.book.....Z}). In these systems the effects 
of environmental perturbation such as tidal interactions and ram pressure stripping are minimized.
The A$_{flux~ratio}$ distribution of our sample is well described by a half-Gaussian function with a  
1$\sigma$ width of 0.13 (possibly smaller $\sigma$ $\sim$ 0.11 if artificially induced effects are reduced). 
Only 9\% of the galaxies show $A_{flux~ratio}$ $>$ 1.26 (2$\sigma$) and only 5\% $A_{flux~ratio}$ $>$ 1.39 (3$\sigma$). 
If it is reasonable to assume that the distribution of intrinsic asymmetries should show a Gaussian distribution, then 
that of Figure~\ref{fig:GaussFit}  is as close as we have ever come to isolating that intrinsic distribution. Comparison with field 
samples clearly shows that effects of nurture result in an excess population of high $A_{flux~ratio}$ values. This excess population 
measured as a deviation from the best Gauss fit is negligible in our sample.   
In isolation it is apparently very unlikely to find galaxies with \hi\ disks showing $A_{flux~ratio}$ $>$ 1.39. 
The small number of such extreme asymmetric profiles found in our sample show double-peaked profiles 
with unequal horns. This is in most cases not caused by contamination by gas-rich companions with systemic velocity 
suitable to create a false or amplified horn. We almost never observe such components in the middle or close to the 
edges (thus broadening one of the horns in) the \hi\ profile. We are unlikely to find a narrower asymmetry distribution 
in any galaxy sample. 

Unfortunately, using \hi\ profiles does not allow us to distinguish between the roles of geometry and kinematics 
in producing an asymmetry. Aperture synthesis maps suggest that asymmetry is usually the signature of kinematic lopsidedness
although galaxies with lopsided HI distributions are not unknown. The rotation curve on one side of the galaxy is usually
steeper than the opposite one. \citet{1999MNRAS.304..330S}  estimated from previous \hi\ maps \citep{1994AAS..107..129B,1996AAS..115..407R,1997PhDT........13V} that the fraction of kinematically
lopsided galaxies may be as large as 15--50\%. It is likely that the few asymmetric \hi\ disks in our sample present this kinematically lopsidedness. However, to make a quantitative relation, it would be necessary to calibrate (statistically) how asymmetry parameters in 2D maps relate to those 1D parameters using single-dish data.



We searched for correlations between internal properties of galaxies and
the measured $A_{flux~ratio}$  parameter. Although we maximized our
sensitivity to internal correlations by removing all galaxies likely to
have been affected by external perturbers, we did not find any strong
correlations between stellar properties and \hi\ asymmetry. We find a
weak correlation between spiral arm strength and \hi\ asymmetry parameter,
in the sense that arms are stronger for galaxies with more symmetric \hi\
profiles. The simplest  interpretation for such a trend would be that gas
asymmetries are more efficiently suppressed by the stronger spiral arm
gravitational torques in more massive galaxies. A connection to spiral arm
strength rather than to bar features should arise
from the larger scale of the former. It has long been known that more
developed spiral arms seem to exist in more luminous galaxies
\citep{1960ApJ...131..558V}. Interestingly, the morphology of spiral arms
is found to depend primarily on parent galaxy properties rather than on
the environment  \citep{2002AJ....124..786V}. Therefore we would expect an
anti-correlation between \hi\ asymmetry and luminosity. However,
Fig.~\ref{fig:afluxratiomorphologyb} shows a hint of the opposite trend.
This might be because unlike van den Bergh's (1960) study, our
study focuses on a sample with a small range in luminosity, as shown in Figure~\ref{fig:characterization}c.

%

The lack of a strong correlation between $L_{FIR}$  and $A_{flux~ratio}$ indicates that the bulk of the star formation and the symmetry of the gaseous disk are not strongly linked, i.e., that induced SF caused by possible interactions in this sample is small compared to that from secular evolution.  Still, there is an excess of about 10\% of asymmetric profiles for the most IR luminous (10~$<$~log($L_{FIR}$[L$_\odot$])~$<$~11) galaxies. This might be linked to recent accretion events in a small number of CIG galaxies. 

In general the asymmetry distribution will likely deviate from a half-Gaussian curve for other samples containing  galaxies that are perturbed by the environment. These samples are the rule while very isolated galaxies are the exception. The intrinsic asymmetry 
distribution found in our sample of isolated galaxies will be skewed toward higher values as a result of these interactions.
This is confirmed by the wider distributions found in samples of field galaxies (Sect.~\ref{sub:samples}) where a higher degree of interaction is expected, given the lack of a strict isolation criterion. 
The deviation from a half-Gaussian curve for the sample in \citet{2005A&A...438..507B} is apparent as shown in Sect.~\ref{sub:samples}, and the distribution is the widest ($\sigma$ = 0.23) among those studied.
Although it has been known for a long time that interacting galaxies usually show larger \hi\ asymmetries \citep{1983AJ.....88..489S}, a statistical analysis using a common \hi\ profile asymmetry parameter in large and well-characterized samples in dense environments is still needed. 

 \citet{2008MNRAS.388..697M} estimate from the density of (collisional) ring galaxies in the local Universe that major asymmetries for $\sim$ 10\% of the galaxies may be produced as a result of a recent fly-by, resulting in a lopsidedness visible over a time scale of 1~Gyr.  Within the uncertainties of this estimate, this may well match the 10--20\% difference between our sample and field samples. 
As reviewed in Sect.~\ref{sec:introduction}, many mechanisms have been proposed other than tidal interactions or ram pressure from the intergalactic medium that may contribute to the asymmetry parameter distribution, such as  minor interactions/mergers (e.g., \citealt{1997ApJ...477..118Z}), gas accretion along cosmological filaments \citep{2005A&A...438..507B}, halo-disk misalignment  \citep{1998ApJ...496L..13L,2001MNRAS.328.1064N}, internal perturbations including sustained long-lived lopsidedness owing to non-circular motions \citep{1980MNRAS.193..313B} or global m=1 instabilities \citep{2007MNRAS.382..419S}. 
In principle the latter physical processes are likely to occur homogeneously for any sample independently of its environmental properties. Thus, the intrinsic asymmetry distribution found in our sample of isolated galaxies is likely due to a combination of these processes. Owing to the lack of spatial resolution in our data, at this moment we cannot distinguish their respective importance. High-resolution observations of isolated galaxies are a good probe to shed light into the origin of these more subtle asymmetries, though they are likely widespread in all kinds of environments.

 We have started a follow-up study of the origin of \hi\ asymmetries using Very Large Array (VLA), expanded VLA and Giant Meter Radio Telescope (GMRT) aperture synthesis \hi\ observations of a subsample of  $\sim$ 20 isolated galaxies, which will be presented in a forthcoming paper. We selected galaxies covering the wide range of asymmetries found in our sample.
One of the isolated galaxies presenting an asymmetric profile in this subsample is CIG~96 (NGC~864), whose \hi\ synthesis imaging from the VLA has been studied in detail in \citet{2005A&A...442..455E}. The asymmetry in the \hi\ profile is associated with a strong kinematical perturbation in the gaseous disk of the galaxy, where on one side the decay of the rotation curve is faster than Keplerian. Although a companion is detected, no tidal tail is found, and it is probably not massive enough to have caused this perturbation. Probably we are witnessing the recent merger of a small gaseous companion.

\section{Summary and conclusions}

We used \hi\ global velocity profiles for a large sample of  isolated galaxies to $i)$ quantify the rate and amplitude of \hi\ asymmetries in isolated spiral galaxies, where environmental processes such as tidal interactions and ram pressure are minimized, $ii)$ study the role of the environment on the \hi\ asymmetries, and $iii)$ study possible correlations between \hi\ lopsidedness and the properties of the stellar component, including their morphological types, signatures of optical perturbation, bar and spiral strengths, as well as optical and FIR luminosities.  
 
To quantify the \hi\ asymmetry, we calculated a flux ratio asymmetry parameter ($A_{flux~ratio}$). 
We restricted our study to a sample of $N$ = 166 galaxies (the \hi\ refined subsample) for which we minimized undesired artificially induced lopsidedness by avoiding large uncertainties owing to the rms of the profile, determination of the mean velocity, and pointing offsets. 

We found that a half-Gaussian curve properly fits the $A_{flux~ratio}$ distribution of this refined sample, with a $\sigma$ = 0.13. We suggest that if we deconvolve other sources of errors such as baseline fitting and random pointing offsets, then the underlying $\sigma$ is reduced  to $\sigma$ $\sim$ 0.11. 
We confirm that by using this sample we effectively minimize nurture effects, because there is a lack of correlation between  \hi\ asymmetries and isolation parameters such as tidal force (one-on-one interactions) and number density.  

We compared the distribution of \hi\ asymmetries of previously studied field galaxies with that of our isolated galaxy sample. A half-Gaussian fit does not successfully reproduce in general the asymmetry distribution of field samples. Indeed, the intrinsic $\sigma$ is larger in field samples than in isolated galaxies. This is likely a result of the lack of an isolation criterion in the selection of field galaxies, which are likely contaminated by interacting objects.
This suggests that environmental mechanisms (producing short-lived effects $\sim$ 1~Gyr) are fundamental mechanisms to produce
\hi\ asymmetries in galaxies, and are indeed responsible for the $\sim$ 10--20\% difference in the \hi\ asymmetry rate we see in field galaxies with respect to isolated galaxies. The asymmetry distribution of galaxies in denser environments is likely even wider and more skewed.

Within the isolated galaxy sample, we did not find any strong correlation between the \hi\ asymmetry and internal properties such as the morphological type, optical and FIR luminosities or signature of interaction. A signature of perturbed optical emission is not a necessary condition for the \hi\  profile to be asymmetric, and vice versa. 
 We found a trend for larger \hi\ asymmetries to be located in more FIR luminous galaxies that are likely interacting objects. We also found evidence that galaxies with higher spiral arm strength have lower \hi\ asymmetries.

 The here presented \hi\ refined subsample can be used to study the origin of intrinsic \hi\ asymmetries in isolated galaxies, and it is also a baseline for samples of galaxies in denser environments with \hi\ data that are properly evaluated for instrumental effects. This can help to shed light into the relative importance of different environmental and internally generated processes in shaping the \hi\ disks of galaxies.

\begin{acknowledgements}
We thank the anonymous referee for a careful reading and very detailed report, which helped to improve this paper significantly.
DE thanks U. Lisenfeld, E. Battaner, P. Vilchez, R. Garrido, and E. Perez for useful comments.
We appreciate the help of the staff members of the different telescopes that have made this work possible (Arecibo, Effelsberg, Nan\c{c}ay and GBT). We thank Francoise Combes for providing the \hi\ asymmetry parameter list in \citet{2005A&A...438..507B}.
DE has been supported by a Marie Curie International Fellowship (MOIF-CT-2006-40298) within the 6$\rm ^{th}$ European Community Framework Programme.
DE, JSM, LVM and SV are supported by DGI Grant  AYA 2008-06181-C02 and the Junta
de Andaluc\'ia (Spain)  P08-FQM-4205. This research has made use of the NASA/IPAC Extragalactic Database (NED) which is operated by the Jet Propulsion Laboratory, California Institute of Technology, under contract with the National Aeronautics and Space Administration. We acknowledge the usage of the HyperLeda database (http://leda.univ-lyon1.fr).
\end{acknowledgements}

\bibliography{AMIGAXI-12272008}

\begin{table*}
\caption{Asymmetry quantification for the \hi\ sample}

 \begin{center}
 \begin{tabular}{rcccccc}
\hline
CIG & Visual classification &  $A_{flux~ratio}$ &$\Delta A (rms)$  & $\Delta A (mean~vel.)$ &$\Delta A_{flux~ratio}$  \\  \hline\hline
2   & 1 & 1.009 & 0.065 & 0.029 & 0.071 \\
4   & 1 & 1.079 & 0.026 & 0.015 & 0.030 \\ 
8   & 0 & 1.080 & 0.028 & 0.018 & 0.034 \\ 
9   & 0 & 1.108 & 0.027 & 0.016 & 0.032  \\
11 & 2 & 1.456 & 0.058 & 0.037 & 0.069 \\ 
...&...&...&...&...&... \\\hline 
\end{tabular}
\end{center}
Note. 1) CIG number, 2) visual classification ($0$ = symmetric, $1$= slightly asymmetric, $2$ = asymmetric), 3) flux ratio asymmetry parameter $A_{flux~ratio}$, 4) $\Delta A (rms)$: uncertainty owing to the rms of the \hi\ profile, 5) $\Delta A (mean~vel.)$: $A_{flux~ratio}$ uncertainty owing to the determination of the mean velocity, and 6) $\Delta A_{flux~ratio}$, the final derived uncertainty of $A_{flux~ratio}$, including the effect of pointing offsets. 
The full list is available in electronic form at the CDS via anonymous ftp to {\tt cdsarc.u-strasbg.fr (130.79.128.5)}, via {\tt http://cdsweb.u-strasbg.fr/cgi-bin/qcat?J/A+A/vvv/ppp} or from {\tt http://amiga.iaa.es}.

\label{tab:tests-coeff}
\end{table*}

\begin{table*}
\caption{ Comparison between the visual classification and $A_{flux~ratio}$}
\begin{center}
\begin{tabular}{ccccc}
\hline
   Visual classification          &  $N$   & Mean & Median & $\sigma$  \\\hline\hline
 Symmetric            & 141   &    1.08     &  1.07 &   0.07 \\
 Slightly asymmetric  & 126  &    1.13      &  1.13 &   0.09 \\
 Asymmetric           & 45   &    1.37      &  1.32 &   0.26 \\\hline
\end{tabular}
\end{center}
\label{tab:lops-visual-sample-histo-symmetry}  
\end{table*}

\begin{table*}
\caption{Comparison of the half-Gaussian $\sigma$ and \hi\ asymmetry rate between samples of isolated/field galaxies
  \label{tab:afluxratio}  }
\begin{center}
\begin{tabular}{lccc}
\hline
Sample                                                    & $N$   & $\sigma$ &   $A_{flux~ratio}$$>$ 1.26\\\hline\hline
 \hi\ refined subsample                     &   166      & 0.13&       9 \%    \\\hline

Haynes et al. (1998)                                      & 104   &   0.13  &         9 \%       \\
Haynes et al. (1998)  no CIGs                       &   80   &     0.13  &   10 \%      \\
Matthews et al. (1998)                                   &  30   &  -   &       17 \%       \\
Bournaud et al. (2005)                                   &  76     & 0.23  &       22 \%     \\\hline
All, no CIGs                                                   & 186   & 0.17 &       16 \%   \\\hline

\end{tabular}
\end{center}
\end{table*}

\clearpage

\end{document}